\documentclass{aa}
\usepackage{natbib,twoopt}
\usepackage{graphicx}
\usepackage{txfonts}
\usepackage{comment}
\usepackage[normalem]{ulem}
\usepackage{xcolor}
\usepackage{tikz}
\newcommand{\bbarolo}[0]{$^{\rm 3D}$BAROLO } 
 
\newcommand{\as}[0]{$^{\prime\prime}$ }

\usepackage[breaklinks=true]{hyperref} 
\bibpunct{(}{)}{;}{a}{}{,}
\makeatletter
  \newcommandtwoopt{\citeads}[3][][]{\href{http://adsabs.harvard.edu/abs/#3}%
    {\def\hyper@linkstart##1##2{}%
     \let\hyper@linkend\@empty\citealp[#1][#2]{#3}}}
  \newcommandtwoopt{\citepads}[3][][]{\href{http://adsabs.harvard.edu/abs/#3}%
    {\def\hyper@linkstart##1##2{}%
     \let\hyper@linkend\@empty\citep[#1][#2]{#3}}}
  \newcommandtwoopt{\citetads}[3][][]{\href{http://adsabs.harvard.edu/abs/#3}%
    {\def\hyper@linkstart##1##2{}%
     \let\hyper@linkend\@empty\citet[#1][#2]{#3}}}
  \newcommandtwoopt{\citeyearads}[3][][]%
    {\href{http://adsabs.harvard.edu/abs/#3}
    {\def\hyper@linkstart##1##2{}%
     \let\hyper@linkend\@empty\citeyear[#1][#2]{#3}}}
\makeatother

\begin{document} 
   \title{The extended molecular gas of the Circinus galaxy and NGC 1097 as seen by APEX}

   \author{Akhil Lasrado
          \inst{1}
          \and
          Claudia Cicone\inst{1}
          \and
          Axel Weiss\inst{2}
          }
          
   \institute{\inst{1}Institute of Theoretical Astrophysics, University of Oslo, P.O. Box 1029, Blindern, 0315 Oslo, Norway\\
            \inst{2} Max-Planck-Institut für Radioastronomie, Auf dem Hügel 69 D-53121 Bonn, Germany \\
              \email{akhil.lasrado@astro.uio.no} \\
             }

   \date{Received March 6, 2025; accepted August 9, 2025}

  \abstract
   {The outer region of the interstellar medium (ISM) is often witness to dynamically important events in a galaxy's evolutionary history such as outflows, inflows, tidal interactions, and mergers, as well as dynamical structures affecting its current evolution such as large-scale bars and spiral arms. Studying the imprints of these processes in the diffuse, extended molecular gas is best achieved by a single dish telescope which can cover a large field of view with good sensitivity to large-scale structures. In this work we present results from Atacama Pathfinder EXperiment (APEX) line emission maps of two nearby galaxies: the Circinus galaxy in the CO(3--2) transition, and NGC~1097 in CO(2--1), covering their full optical extents. We detect molecular gas at the largest extents seen for these galaxies yet, at up to 5$^{\prime}$ ($r \approx 6$ kpc) for the Circinus galaxy, and $4^{\prime}.5$ ($r \approx 18$ kpc) for NGC~1097, and compute total CO luminosities of $L^{\prime}_{\mathrm{CO(3-2)}} =  (1.5\pm0.4)\times10^{8}$ K km s$^{-1}$ pc$^{2}$ and $L^{\prime}_{\mathrm{CO(2-1)}} = (7.0\pm1.7)\times10^{8}$ K km s$^{-1}$ pc$^{2}$, corresponding to molecular gas masses $(2.1\pm1.0)\times10^{9}$ M$_{\odot}$ and $(4.7\pm1.9)\times10^{9}$ M$_{\odot}$, respectively. We further analyze the large-scale gas kinematics through position-velocity diagrams and 3D tilted ring modeling using the \bbarolo code. We detect notable features in both galaxies beyond their well-studied bright central regions: in the Circinus galaxy we detect molecular gas embedded in a bar-like structure, whose kinematic signature is also evident in the major axis position-velocity diagram, and in NGC~1097, we observe tidal molecular gas structures involved in the interaction of NGC~1097 with the companion galaxy NGC~1097A. The clear detection of such structures in the molecular gas shows promise in conducting large-scale molecular gas studies toward nearby galaxies with APEX and, in the future, the Atacama Large Aperture Submillimeter Telescope (AtLAST).}


   \maketitle

\section{Introduction}

The cold molecular phase of the interstellar medium (ISM) in galaxies is an essential ingredient for the formation of stars, and obtaining the properties of this gas provides key constraints about the efficiency and mechanisms involved in the star formation process. On galaxy scales, the amount and physical conditions of molecular gas govern the evolution of a galaxy and its position on the star-forming main sequence \citep{2007ApJ...670..156D, 2007A&A...468...33E, 2022ARA&A..60..319S}. In regular spiral galaxies, molecular gas is generally co-spatial with the optical disk and located in the spiral arms. The nuclear regions can host higher concentrations of molecular gas, especially in starburst  \citep{1991mss..book..233S, 1995A&A...300..369A, 2009A&A...506..689I} and barred galaxies \citep{1999ApJ...525..691S, 2005ApJ...630..837J, 2022A&A...666A.175Y}.

On the outskirts of galaxies, the molecular ISM component can be more diffuse and therefore harder to detect. Nevertheless, gas at the outskirts of galaxies is a key participant of the baryon cycle and can be disturbed and redistributed due to tidal interactions with companion galaxies or due to mergers \citep{2001A&A...372L..29A, 2022ApJ...936L..11S}, and can also be transported from 10s of kpc scales to active centers of galaxies at $\sim1$ kpc scales thanks to inflows driven by non-axisymmetric potentials. It is also useful to study the conditions for the formation of molecular gas and star formation activity in low density environments, since these can differ greatly in metallicity, stellar radiation field, kinetic temperatures, and mechanisms that stimulate the neutral hydrogen (HI) to H$_{2}$ phase transition compared to star-forming regions in the Milky Way or inner disks of galaxies \citep{2017ASSL..434.....K, 2017ASSL..434..175W, 2024MNRAS.528.4746U}.

Nearby galaxies provide the opportunity to map molecular gas at resolutions that allow us to link the physical properties of the gas and the processes they are involved in, providing a link between galactic studies in the Milky Way, where the extended molecular ISM is prohibitive to observe due to its large angular extent in the sky, and higher redshift sources, where extended molecular ISM reservoirs - out to the circumgalactic medium (CGM) - are easier to study thanks to the higher angular distance. 

Interferometers are not well suited for mapping extended emission in nearby objects, as they have limited sensitivity on areas larger than their synthesized beam \citep{2024A&A...688A..30B} and miss flux on scales larger than their maximum recoverable angular scale. In contrast, single-dish telescopes are ideal for this purpose, offering high sensitivity to large-scale structures. There have been previous surveys mapping the large scale molecular gas in local galaxies such as HERACLES \citep{2009AJ....137.4670L}, VERTICO \citep{2021ApJS..257...21B}, ATLAS$^{3D}$ \citep{2013MNRAS.432.1796A}, COMING \citep{2019PASJ...71S..14S}, but there are few large-scale maps of very nearby galaxies with newer facilities \citep{2004ApJ...612..860M, 2004A&A...413..505L, 2022A&A...662A..89D}. One of the foremost single dish observatories is the Atacama Pathfinder Experiment (APEX) telescope. APEX has successfully been used in studies involving the large scale molecular gas of nearby galaxies, such as to study extended diffuse molecular gas in ram pressure stripped tails \citep{2014ApJ...792...11J}, to obtain the total molecular gas reservoir of dwarf galaxies \citep{2014A&A...564A.121C}, and also to produce maps of the Large and Small Magellanic Clouds \citep{2024A&A...682A.137G} for studying feedback. There are however few studies focusing on large scale mapping of molecular lines in nearby galaxies.

In this paper, we present results from APEX observations of two nearby galaxies encompassing their full optical extents, the Circinus galaxy (henceforth Circinus), and NGC 1097. Both targets are well studied and have an immense archive of multi-wavelength observations, including of the large-scale molecular gas studies in \citet{2008MNRAS.389...63C} and \citet{2001PhDT.......237C} for Circinus and NGC 1097, respectively. However our data for Circinus covers a larger, more regular area than the data in \citet{2008MNRAS.389...63C}, and our data for NGC~1097 is more sensitive than the data in \citet{2001PhDT.......237C}, and therefore our maps provide the most updated view of the large-scale molecular gas in these galaxies. An unbiased map at these extents also helps constrain the total molecular gas content of the observed galaxies without selecting for regions with far-infrared, HI, or H$\alpha$ emission, which may not perfectly trace CO emission \citep{2010A&A...520A.107B, 2017ASSL..434..175W}. This is, to our knowledge, the first such study performed with the APEX telescope, and can serve as a pathfinder study for potential future single dish sub-millimeter telescopes such as the Atacama Large Aperture Submillimeter Telescope (AtLAST)\footnote{\href{www.atlast.uio.no}{www.atlast.uio.no}}.

The Circinus galaxy is a nearby, highly inclined spiral (SAb) galaxy \citep{1977A&A....55..445F, 1991rc3..book.....D} hosting the nearest Type-2 Seyfert nucleus \citep{1994Msngr..78...20M, 1996A&A...315L.109M}, at a distance of 4.2$\pm$0.8 Mpc (20.4 pc per arcsecond) \citep{2009AJ....138..323T}. Circinus also hosts a galactic outflow seen in the blueshifted ionized gas along the minor axis up to 30$^{\prime\prime}$ ($\sim 600$ pc) \citep{1994Msngr..78...20M, 1997ApJ...479L.105V, 2000AJ....120.1325W}. The outflow is also detected in the radio continuum by \cite{1998MNRAS.297.1202E}, and also in molecular gas features within the central kpc \citep{2016ApJ...832..142Z}. The atomic and molecular large scale gas of Circinus has been studied in some detail. The atomic gas was first mapped out in low resolution by the Parkes radio telescope, revealing a massive hydrogen disk larger than 1$^{\circ}$ (70 kpc) in diameter \citep{1977A&A....55..445F}, while higher resolution maps of the distribution and kinematics of HI by \cite{1999MNRAS.302..649J} and \cite{2008MNRAS.389...63C} revealed a kinematically warped disk, and an atomic ring of radius $\sim$10 kpc hosting a bar that terminates at the center in a smaller 1 kpc ring. The surface brightness of the molecular medium in Circinus is dominated by emission from the molecular ring of radius $\sim$600 pc, and this central gas-rich region has been the main focus of studies probing isotopologs of CO and high density gas tracers \citep{1991A&A...249..323A, 1998A&A...338..863C, 2001A&A...367..457C, 1994Msngr..78...20M, 2014A&A...568A.122Z, 2016ApJ...832..142Z, 2023Sci...379.1323E}. The large-scale molecular gas has been studied in detail in \cite{2008MNRAS.389...63C}, which also shows signs of warping (coinciding with the positions of the bar seen in the atomic gas), detection of CO at 3 kpc, and a higher fraction of molecular to atomic gas within the molecular ring.

NGC 1097 is a prototypical barred spiral (SBb) galaxy \citep{1987A&A...172...32H, 1995AJ....110.1009B} at a distance of 13.58$\pm$2.05 Mpc (65.7 pc per arcsec) \citep{2021MNRAS.501.3621A}. The galaxy shows two prominent dust lanes that terminate in a central starburst ring ($r\sim700$ pc) which hosts high star formation rates (SFR $\simeq$ 2 M$_{\odot}$ yr$^{-1}$) \citep{2010A&A...518L..59S, 2011ApJ...736..129H, 2021ApJ...923..150L}, and within which lies a r$\sim$300 pc circum-nuclear disk around the central low luminosity active galactic nucleus (AGN) \citep{2006ApJ...643..652N}. Toward the northwest of the galaxy lies NGC~1097A, an elliptical companion galaxy, which is interacting with NGC~1097 as seen from the disturbed outer spiral arms of the latter. The large scale atomic gas of NGC 1097 was first mapped out by \cite{1989ApJ...342...39O}, and most recently by \cite{2003ApJ...585..281H}, covering the bar and outer spiral arm pattern, and no HI gas was detected at the position of NGC~1097A in either observation. NGC~1097 also hosts four faint optical jets, which are primarily comprised of stars with no atomic gas detected by \cite{2003ApJ...585..281H}, who posit a minor merger origin to the jets. Large scale CO(1--0) observations of NGC~1097 were obtained by \cite{2001PhDT.......237C} covering its optical extent, who primarily find molecular gas in the nuclear region and bar, as well as in the inter-arm region where a deficit in the HI gas is seen. Higher resolution molecular gas observations of NGC~1097 were obtained as part of the PHANGS-ALMA program \citep{2021ApJS..257...43L, 2021ApJS..255...19L}. This data was used by \cite{2023MNRAS.523.2918S} to find the inflow rate along the bar to be $\dot{\text{M}} \simeq (3.0\pm2.1)$ M$_{\odot}$ yr$^{-1}$, which is comparable to the SFR of the nuclear ring, suggesting that star formation in the ring might be regulated by the radial flow of gas through the bar.

Both targets are highly star-forming, gas-rich galaxies, and are positioned at the upper end of the star-forming main sequence \citep{1998ApJ...498..541K,2022ARA&A..60..319S,2015ApJ...801L..29R, 2024A&A...687A.244H}, and show dynamically prominent features such as warps, bars, and tidal tails, and therefore make them suitable targets for studying their large-scale molecular gas. Both of these galaxies are also nearby and do not have recent large-scale single dish maps of their molecular gas content. Complete physical properties of both galaxies are listed in Table~\ref{tab:galprops}. The paper is organized as follows: In Sect. \ref{sec:observ} we present details of the observations, the observing strategy, and reduction methods. In Sect. \ref{sec:methodology} we describe the methodology for analyzing the APEX data. Then the results of our analysis are presented: for Circinus in Sect. \ref{sec:circ_results}, and for NGC~1097 in Sect. \ref{sec:ngc1097_results}. In Sect. \ref{sec:discuss}, we discuss notable features and their implications. Our conclusions are summarized in Sect. \ref{sec:conc}.

\begin{table*}
\centering
\fontsize{7.7pt}{7pt}\selectfont
\caption{Details of the APEX observations.}
\label{tab:obs_dets}
\begin{tabular}{lcccccccc}
\hline
\hline\\[-1mm]
        Galaxy & Field of view & Line & Instrument & Observing Dates & $\theta_{\rm FWHM}$ & On-source integration time & Jy/K & RMS ($\Delta v = 5$ km/s)\\
         & (arcmin$^{2}$) & & & & ($^{\prime\prime}$) & (hours) & & $T^{*}_{A}$ (mK) \\[1mm]
\hline \\ [-1mm]
Circinus & $10'\times10'$ & CO(3--2) & LAsMA     & 7 Jul 2023 - 6 Aug 2023 & 19.1       & 8.3             & 37 & $\sim 30$ \\[1mm]
NGC 1097  & $12'\times12'$ & CO(2--1) & nFLASH230 & 1 Sept 2023 - 20 Nov 2023 & 28.7       & 11.9             & 35 & $\sim 10$ \\[1mm]
\hline
\end{tabular}
\end{table*}

\section{Observations}\label{sec:observ}
\subsection{Observing strategy}
The observations were performed by the Atacama Pathfinder EXperiment (APEX), a 12-meter single dish sub-millimeter telescope located on the Chajnantor Plateau in Chile \citep{2006A&A...454L..13G}. We observed Circinus and NGC 1097 in the CO(3--2) and CO(2--1) emission lines, respectively, which enable us to capture the bulk of the molecular gas at sub-kpc resolution in Circinus, and at $\sim2$ kpc resolution in NGC~1097. The initial goal was to map both galaxies in the CO(3--2) line to achieve a compromise between resolution and sensitivity; however, due to the unavailability of the LAsMA instrument at the time, NGC~1097 was instead mapped in CO(2--1) with nFLASH230.

\subsection{Data reduction}\label{sec:datred}
Data reduction was performed using \textsc{GILDAS/CLASS} followed by further masking in \texttt{python}. All quoted velocities are redshift-corrected. Table \ref{tab:obs_dets} summarizes key information of the observations. The reduction methodology in \texttt{CLASS} is described in detail for both galaxies below, while the \texttt{python} masking is described in Sect. \ref{sec:masking}.

\subsubsection{Circinus}
Observations of the CO(3--2) line ($\nu_{\text{obs}} = 345.2960$ GHz) in Circinus were obtained with the 7-pixel receiver LAsMA (large APEX sub-millimeter array) in a period between 7 July and 6 August 2023, with an on-source observing time of 8.3 hours, with an average precipitable water vapor (PWV) of 0.93 mm ($\sigma$ = 0.2 mm). The maps were observed in total power on-the-fly (OTF) mode in a rectangular pattern and cover a field of view (FoV) of $10^{\prime} \times 10^{\prime}$. The data were reduced using a \texttt{CLASS} script, in which we subtract a first order baseline from the spectra (which have units of $T_{A}^{*}$) scan-by-scan after selecting a bandwidth of 2.3 GHz (2000 km/s) centered on the rest frequency of CO(3--2) ($\nu_{\text{rest}} = 345.796$ GHz), and masking the line emission in a window in the range $v \in (-500, 500)$ km/s, and re-bin the spectra from a native spectral resolution of 0.43 km/s to a resolution of 5 km/s. For these observations, scans from the receiver backend APL-309-F402 (retaining APL-308-F402 to APL-313-F402) were dropped due to baseline instabilities (comprising $\sim7\%$ of all scans). To convert the cube to the main beam brightness temperature scale, we used a main beam efficiency of $\eta_{\text{MB}} = 0.76\pm0.06$ for August 2023, listed on the APEX telescope efficiencies page\footnote{\href{www.apextelescope.org/telescope/efficiency/}{www.apextelescope.org/telescope/efficiency/}}, and for our analysis use a Kelvin to Jansky conversion factor of 37 Jy/K. We further cropped the data cube to a velocity range of -370 to 500 km/s to exclude galactic CO(3--2) emission which manifests itself at $v\sim-470$ km/s in the Circinus cube. The final cube has a median rms value of $\sim30$ mK in $T_{A}^{*}$ units ($\sim1.1$ Jy/beam), and has a beam full width at half maximum (FWHM) at 345.8 GHz of 19.1$^{\prime\prime}$ (380~pc at the distance of Circinus).

\subsubsection{NGC 1097}
CO(2--1) observations of NGC 1097  ($\nu_{\text{obs}} = 229.5601$ GHz) were performed with the nFLASH230 channel of the nFLASH (new FaciLity APEX Submillimeter Heterodyne) instrument between 1 September and 20 November 2023, with an on-source time of 11.9 hours, with average PWV values of 1.98 mm ($\sigma$ = 0.81 mm). The observations were carried out in total power OTF mode in a rectangular pattern. The FoV of the observations cover a $12^{\prime} \times 12^{\prime}$ region. We used \texttt{CLASS} for the data reduction, subtracting a first order baseline from the spectra (in $T_{A}^{*}$) scan-by-scan. For the baseline subtraction, we selected a bandwidth of 1.9 GHz (2500 km/s) centered on the CO(2--1) rest frequency ($\nu_{\text{rest}} = 230.538$ GHz), and used a mask from $v \in (-500,500)$ km/s to avoid line emission. The spectra were then re-binned from a native spectral resolution of 0.33 km/s to a resolution of 5 km/s. The main beam efficiency for converting to the main beam brightness temperature scale was $\eta_{\text{MB}} = 0.81\pm0.07$ (using values calibrated with Jupiter), and a Kelvin to Jansky conversion factor of 35 Jy/K for nFLASH230 was used over the period of observations from the APEX telescope page for our analysis. The cube was further cropped to a velocity range of -485 to 500 km/s to match the velocity range of the PHANGS-ALMA cube (see below). The median rms noise of the final cube is $\sim10$ mK in $T_{A}^{*}$ units ($\sim0.35$ Jy/beam), with a beam FWHM at 230.538 GHz of 28.7$^{\prime\prime}$ (1.9~kpc at the distance of NGC~1097).

\begin{table*}
\centering
\footnotesize
\caption{Properties of Circinus and NGC 1097.}
\label{tab:galprops}
\begin{tabular}{lccc}
\hline
\hline \\ [-2mm]
        Property & Circinus galaxy & NGC 1097 \\ [1mm]
\hline \\ [-1mm]
$\alpha_{\text{J2000}}$ & $14^{h}13^{m}10^{s}.4$ & $2^{h}46^{m}18^{s}.9$ \\[1mm]
$\delta_{\text{J2000}}$ & $65^{\circ}20^{\prime}20^{\prime\prime}$ & $30^{\circ}16^{\prime}29^{\prime\prime}$\\[1mm]
R$_{25}$                & $6.9^{\prime}\times3.0^{\prime}$ ($8.5\times3.6$ kpc) (1) & $9^{\prime}.3\times6^{\prime}.3$ ($36.8\times24.8$ kpc) (1) \\[1mm]
Luminosity Distance (\textit{D$_{L}$}) & $4.2\pm0.7$ Mpc (2) & $13.58\pm2.05$ Mpc (3) \\[1mm]
Redshift (\textit{z})   & 0.001448 & 0.00424 \\[1mm]
$v_{\text{LSR}}$        & 434 km/s (4) & 1257.5 km/s (5) \\[1mm]
Position Angle (P.A.)   & $210^{\circ}$ (6), (7) & $122^{\circ}.4\pm3.6$ (8)\\[1mm]
Inclination (\textit{i}) & 65$^{\circ}$ (6) & $(48.6\pm6)^{\circ}$ (8)\\[1mm]
$\log_{10}M_{*}$        & 10.98 (9) & 10.76 (5)\\[1mm]
SFR                     & 4.3 M$_{\odot}$ yr$^{-1}$ (9, 10) & 4.78 M$_{\odot}$ yr$^{-1}$ (5)\\[1mm]
AGN Luminosity          & 10$^{10}$ $L_{\odot}$ (11)& $2.2\times10^{8}$ $L_{\odot}$ (12) \\[1mm]
\hline
\end{tabular}
\\ [1mm]
{\raggedright References: (1)  \cite{1991rc3..book.....D}, (2)  \cite{2009AJ....138..323T}, (3)  \cite{2021MNRAS.501.3621A}, (4)  \cite{2016ApJ...832..142Z}, (5)  \cite{2021ApJS..257...43L}, (6) \cite{2018ApJ...867...48I}, (7)  \cite{2008MNRAS.389...63C}, (8)  \cite{2020ApJ...897..122L}, (9)  \cite{2012MNRAS.425.1934F}, (10)  \cite{1996A&A...315L.109M}, (11) \cite{2006ApJ...643..652N}, (12) \cite{2009ApJ...692..556R}.}
\end{table*}
\begin{figure}
\centering
\includegraphics[width=0.49\textwidth]{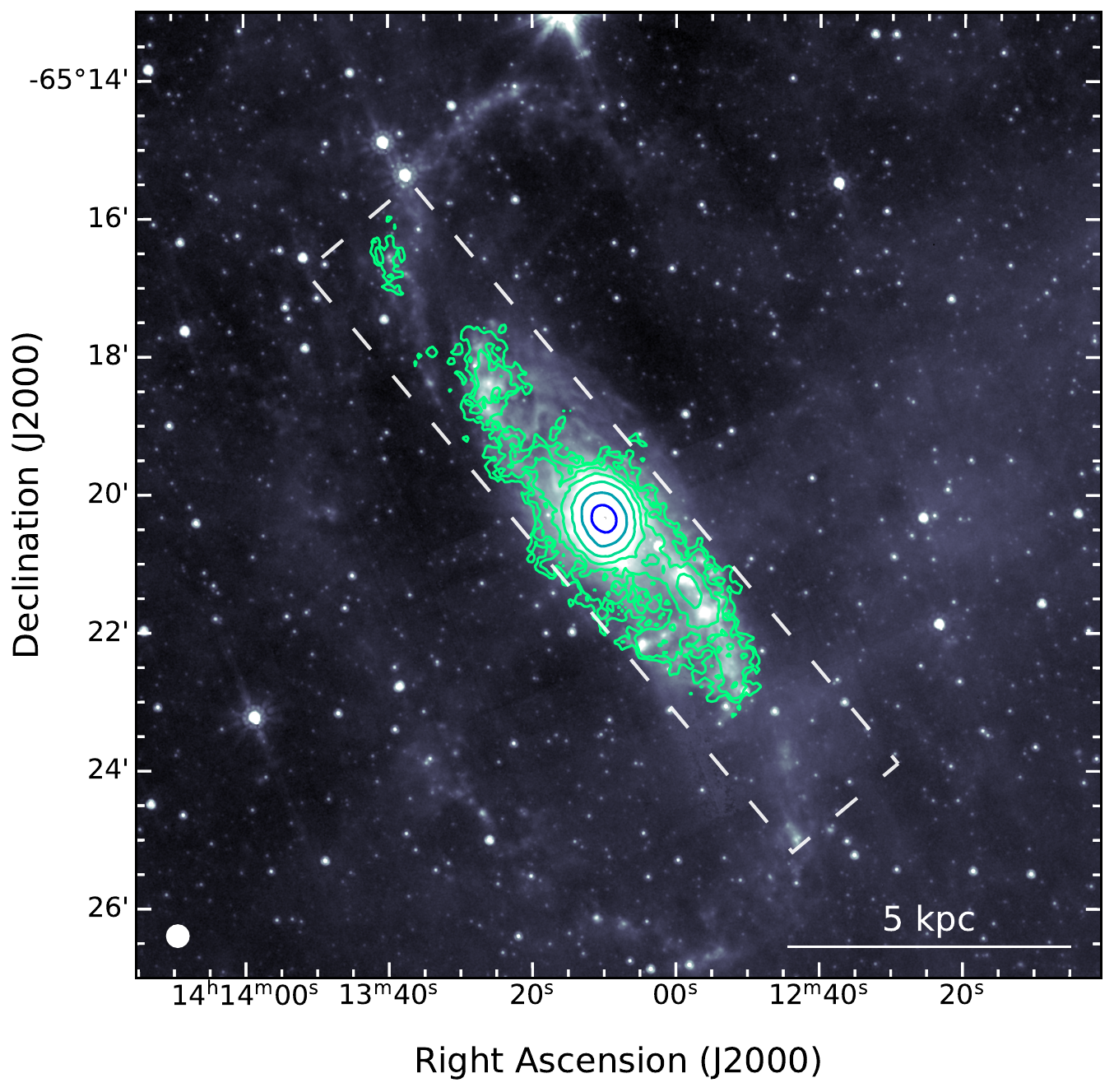}
\caption{CO(3--2) integrated intensity contours computed from -370 to 500 km/s using the dilated masking technique for Circinus, overplotted on the Spitzer/IRAC 8 $\mu$m continuum map. The contours are plotted on 7 log-spaced intervals, ranging from 0.36 Jy km/s to 121 Jy km/s. The dashed box shows the slice used for the major axis position-velocity cut. The beam size of the APEX observations (19$^{\prime\prime}$.1) is shown in the bottom left.}
\label{fig:circinus_f1}
\end{figure}

\begin{figure}
\centering
\includegraphics[width=0.49\textwidth]{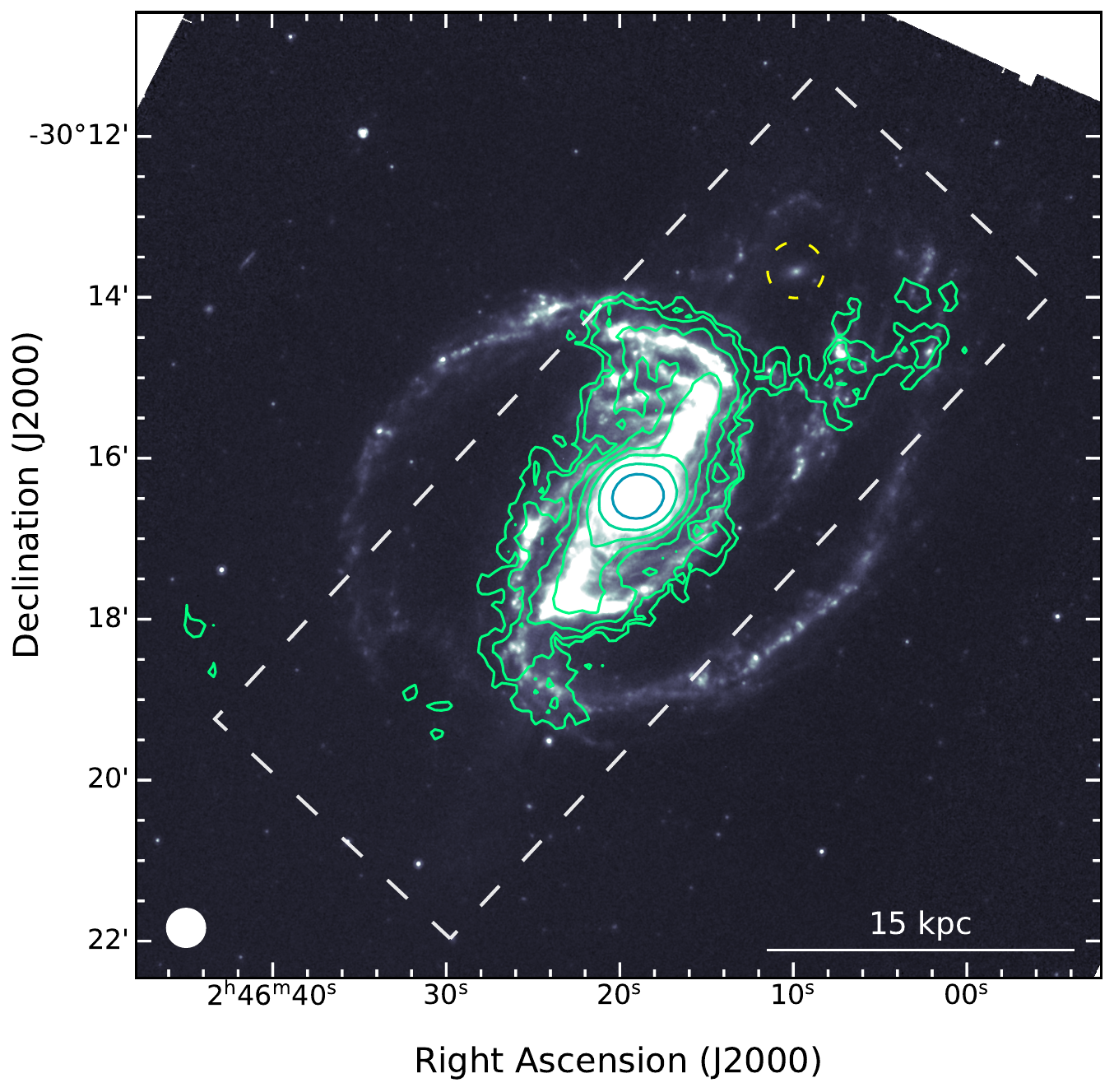}
\caption{CO(2--1) integrated intensity contours calculated from -485 to 500 km/s using the dilated masking technique for NGC 1097 overplotted on the Spitzer/IRAC 8 $\mu$m continuum map. The contours are plotted on 8 log-spaced intervals, ranging from 0.15 Jy km/s to 75.6 Jy km/s. The dashed yellow circle marks the position of the companion galaxy NGC 1097A. The dashed box shows the slice used for the major axis position-velocity cut. The beam size of the APEX observations (28$^{\prime\prime}$.7)  is shown in the bottom left.}
\label{fig:ngc1097_f1}
\end{figure}

\subsection{Archival ALMA data}

We primarily performed comparisons of our data with archival observations from the Atacama Large Millimeter/submillimeter array (ALMA) observatory\footnote{\href{https://www.almascience.nrao.edu/aq/}{www.almascience.nrao.edu/aq/}} for both galaxies. For Circinus, we used high resolution ALMA 12-m array (excluding ACA and TP array data) observations of the CO(1--0) line from \cite{2016ApJ...832..142Z} which have a beam size of $2.7^{\prime\prime} \times 2.0^{\prime\prime}$ and cover the inner 1$^{\prime}$ of the galaxy. For NGC~1097 we used data from the Physics At High Angular Resolutions (PHANGS) project, of the CO(2--1) line from \cite{2021ApJS..257...43L} which have a beam size of $2.0^{\prime\prime}\times2.0^{\prime\prime}$ and cover a rectangular area of $4.8^{\prime}\times2.7^{\prime}$. The PHANGS data is the result of the combination of total power and interferometric data from the 12-m and 7-m arrays using the ``\texttt{feather}'' technique in \texttt{CASA}.

\section{Methodology}\label{sec:methodology}
\subsection{Masking}\label{sec:masking}

The primary goal of our observations is to recover fainter emission. It is therefore important to systematically extract signal from the data through an appropriate masking strategy. To this end we employed a dilated masking technique described in detail in \citet{2024A&A...682A.137G} which is based on the dilated masking approach in \citet{2021MNRAS.502.1218R}. The goal of the masking is to locate regions that despite having a low signal-to-noise (S/N), neighbor high S/N voxels, and are therefore more likely to be real.

For both targets, we adopted values of $R_{1} = 1.5$, $R_{2} = 4.0$, and $R_{3} = 3.0$, using the convention for the S/N parameters from \cite{2024A&A...682A.137G}, which are chosen to maximize the detection of low-surface-brightness features while also trying to exclude false detections. The masked data were then used to compute moment maps (integrated intensity, velocity centroid, and velocity dispersion). Figures \ref{fig:circinus_f1} and \ref{fig:ngc1097_f1} show contours of the integrated intensity map derived from the dilated masking procedure for both galaxies overlaid on the Spitzer 8 $\mu$m continuum map (obtained from the Spitzer Enhanced Imaging Products archive\footnote{\href{https://www.ipac.caltech.edu/doi/irsa/10.26131/IRSA433}{https://www.ipac.caltech.edu/doi/irsa/10.26131/IRSA433}}).

\subsection{CO line luminosity}\label{sec:method_lco}

The line luminosity of the CO lines were obtained from the velocity-integrated fluxes using the prescription given by \citet{2005ARA&A..43..677S}:
\begin{equation}\label{eq:lineluminosity}
    L^{\prime}_{\mathrm{CO}} [\mathrm{K~km~s^{-1}} \mathrm{pc}^{2}] = 3.25 \times 10^{7}\frac{D_{L}^{2}}{\nu_{obs}^{2}(1+z)^{3}}\int_{\Delta v}S_{\nu}~dv,
\end{equation}
where $D_{L}$ is the luminosity distance in Mpc, $z$ is the redshift, $\nu_{obs}$ is the observed line frequency, which is the rest frequency of the line $\nu_{rest}$ divided by ($1+z$), $S_{\nu}$ is the flux density, and $dv$ is the velocity increment.

After converting the CO(3--2) or CO(2--1) luminosity to one for CO(1--0), we derived molecular gas masses using a CO to H$_{2}$ conversion factor $\alpha_{\mathrm{CO}}$,
\begin{equation}\label{eq:alphamass}
    M_{mol} = \alpha_{\mathrm{CO}} L^{\prime}_{\mathrm{CO}}
\end{equation}
In this work we used a constant Milky Way derived $\alpha_{\mathrm{CO}}$ value for the CO(1--0) line of $4.35 \pm 1.3$ $M_{\odot}$ / (K km s$^{-1}$ pc$^{2}$) \citep{2013ARA&A..51..207B} for both galaxies.

\subsection{Tilted ring modeling}\label{sec:method_barolo}

To study the dynamical structure and distribution of the molecular gas in Circinus and NGC~1097, we modeled the data in concentric rings, fitting their geometry and kinematics. We used the widely used tool \bbarolo \citep{2015MNRAS.451.3021D} for this purpose. \bbarolo fits the full position-position-velocity data cube, solving for four geometrical parameters $x_{\rm geom}=(x_{0}, y_{0},\phi,i)$ and four kinematic parameters, $x_{\rm kin} = (V_{sys}, V_{rot}, V_{rad}, \sigma_{gas})$.

Before the fitting, we smoothed the data spatially and spectrally to bring out low-S/N features, and used the dilated masks as input masks to the \texttt{3DFIT} task in $^{\rm 3D}$BAROLO. For both our galaxies, we chose the azimuthal normalization, since this returns better fitting models for the kinematics. For the modeling we used the following methodology, similar to that used in \cite{2021ApJ...923..220D} and \cite{2023A&A...675A..88A}.

First, we used the integrated intensity map and velocity centroid maps to obtain initial guesses for the central position and position angle of the galaxies. The initial guess for the inclination was obtained from higher resolution optical data for both galaxies, and the disk width of the galaxies was set to one pixel for NGC 1097 and two pixels for Circinus, which is highly inclined. We used these parameters and run \bbarolo with a two-stage fitting process, which includes a regularization phase where the position angle and inclination are fit according to bezier curves to avoid artificial discontinuities in the geometry.

While the resulting position angle and inclination so obtained are smooth due to regularization, the kinematics still contain discontinuities which can be attributed to some rings having few data points, high inclination, and poor S/N in some cases. Therefore in the next step, we manually adjusted both $V_{rot}$ and $\sigma_{gas}$ to smoothly vary over all the modeled rings. The $V_{rot}$ was set to follow the rotation curve indicated by the major axis position-velocity (PV) diagram, while $\sigma_{gas}$ was chosen to broadly follow a smooth decline from the center. This was done iteratively, using the manually defined kinematic values as inputs again to \bbarolo with smaller bounds. Once a satisfactory rotation curve and velocity dispersion profile were reached, we ran \bbarolo with this model and calculated the associated errors in the geometry keeping the kinematics fixed. In the final step, we fixed the geometry and kinematics of the disk and include a radial velocity component as a free parameter to the model. We further describe the initial parameters for the modeling of both galaxies below.

For the case of Circinus, we smoothed\footnote{For the smoothing, we use the \texttt{SpectralCube.convolve\_to()} function.} the data to a spatial resolution of 23$^{\prime\prime}$ and velocity resolution of 10 km/s. The center of the galaxy ($\alpha_{J2000} = 14^{h}13^{m}10^{s}.4$, $\delta_{J2000} = 65^{\circ}20^{\prime}20^{\prime\prime}$) was obtained from the integrated intensity map. For the first step the initial guess for the inclination of $65^{\circ}$ was obtained from higher resolution optical data \citep{1977A&A....55..445F, 2000AJ....120.1325W}. The position angle was initially set to the position angle in the central region estimated from iso-velocity contours of the velocity centroid map (205$^{\circ}$), and fit for 27 rings with a width of 10\as each. Given the highly warped nature of the disk of Circinus, we allowed the inclination and position angle to vary by up to 20$^{\circ}$ and 40$^{\circ}$ in each ring for the first step. Finally, the radial velocity was included as a parameter and set to an initial value of 20 km/s.

For NGC~1097, we did not expect the geometry to not vary as much as for Circinus, since NGC~1097 has a mostly regular disk \citep{2003ApJ...585..281H} aside from some signs of interactions toward the northwest. We smoothed the data to a spatial resolution of 33\as and a velocity resolution of 10 km/s. In this case the center was fixed to ($\alpha_{J2000} = 2^{h}46^{m}18^{s}.9$, $\delta_{J2000} = -30^{\circ}16^{\prime}29^{\prime\prime}$), the initial inclination guess of $50^{\circ}$ was obtained from \cite{2020ApJ...897..122L}, who use a photometric prior from high resolution Spitzer/S$^{4}$G data \citep{2010PASP..122.1397S}. Although the position angle from \cite{2020ApJ...897..122L} is $\sim122^{\circ}$, we used a P.A. of 137$^{\circ}$ \citep{1996ApJ...472...83S, 2006ApJ...641L..25F} since it appears as a better match to the P.A. of the disk by eye. We fit 16 rings with a width of 15\as each out to a radius of 232$^{\prime\prime}$. We allowed the inclination and position angles to vary by up to 15$^{\circ}$ and 30$^{\circ}$, respectively in the first step. The dispersion obtained in this step once again showed unphysical spikes in the inner rings, and was therefore adjusted to decline smoothly. Finally for the fit including the radial velocity component an initial value of 20 km/s was chosen. 

\section{Results}\label{results}
\subsection{Circinus}\label{sec:circ_results}
\subsubsection{Gas distribution}\label{sec:circ_gasdist}
\begin{figure}
\centering
\includegraphics[width=0.49\textwidth]{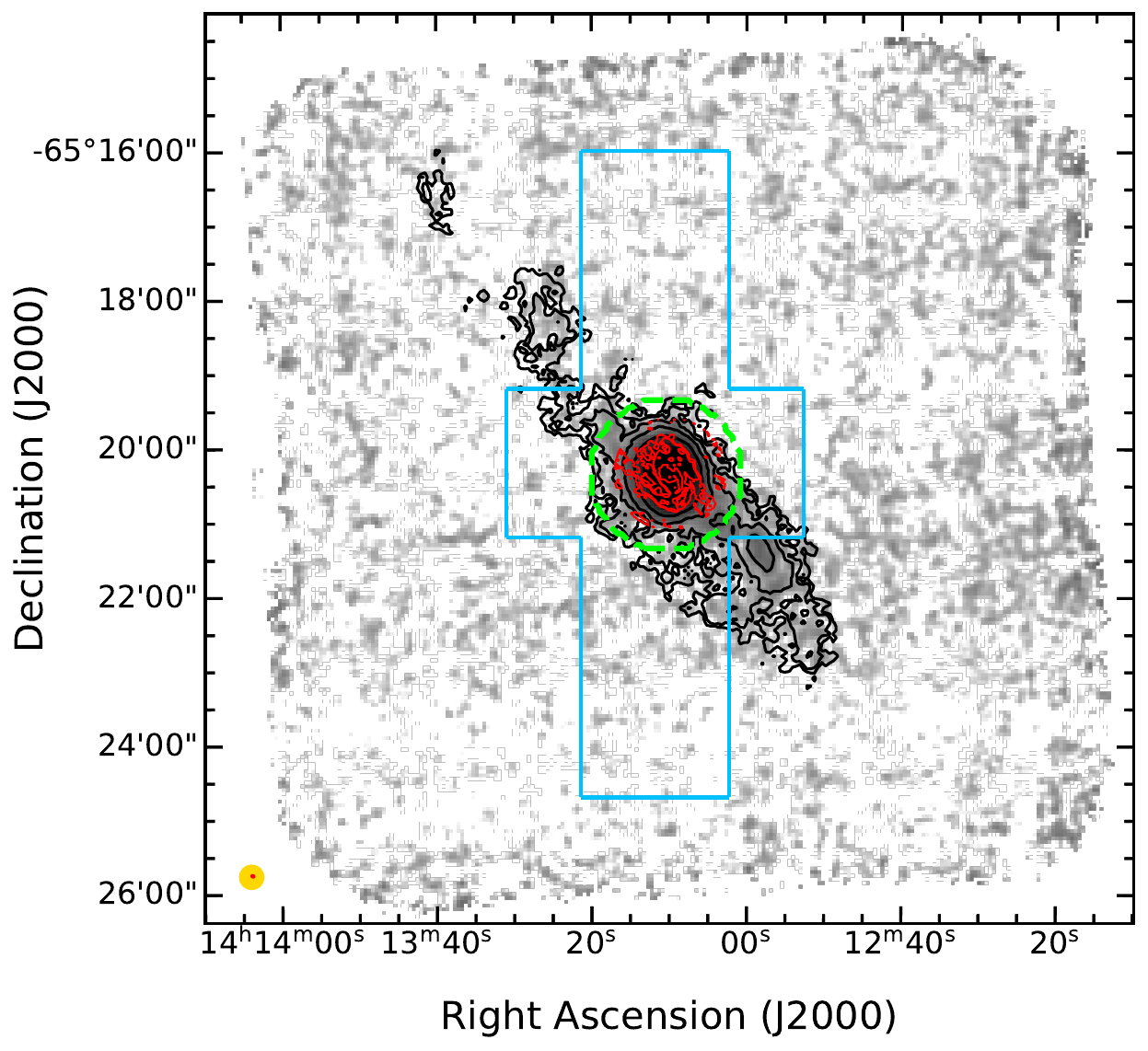}
\caption{CO(3--2) integrated intensity map computed from the unmasked data from $-$220 to 220 km s$^{-1}$ for Circinus. The APEX contours are plotted from the masked integrated intensity map in 7 log-spaced levels from 0.36 Jy km/s to 121 Jy km/s. The red contours show the integrated intensity map of the ALMA 12-m array  CO(1--0) data from \cite{2016ApJ...832..142Z}. The 4 contours are plotted on log-spaced intervals, starting at 1.5 Jy/beam km/s to 63 Jy/beam km/s and the edge of the ALMA map is masked to avoid spurious emission from noise. The blue polygon outlines the footprint of the SEST CO(1--0) map of \cite{2008MNRAS.389...63C}, while the dashed green circle marks the 1$^{\prime}$ (1.2 kpc) radius aperture that encloses the bright central region. The beam size of the APEX and ALMA observations is shown in the bottom left.}
\label{fig:circinus_totalm0}
\end{figure}

To visualize the gas distribution and the extent of the emission detected by APEX, we plot the unmasked integrated intensity map overlaid with the largest extent ALMA map (from \cite{2016ApJ...832..142Z}) in Figure \ref{fig:circinus_totalm0}. In the bright central region, we do not see many features besides just a radial drop-off in the intensity. The furthest gas detection from the center is in the northeast at a distance of $\sim$4.8$^{\prime}$ ($\sim$5.9 kpc), while in the southwest, the furthest detection is at $\sim$3$^{\prime}$ ($\sim$3.6 kpc). The gas distribution in the SEST CO(1--0) integrated intensity map presented in Figure 3 of \cite{2008MNRAS.389...63C} is similar to the APEX observations, especially in CO(1--0) where the southwestern disturbances are more prominent than in CO(2--1). However, due to the larger area mapped with APEX, we detect emission toward the southwest and northeast outside the area mapped by SEST (see Figure \ref{fig:circinus_totalm0}). The low intensity large-scale structures beyond the bright center on both the northeast and southwest appear to be continuous coherent structures connected to the center, and both these structures appear to wind where their emission terminates in a loose spiral pattern. This structure therefore appears to be a molecular counterpart to a bar terminating in spiral arms that is seen in the atomic gas and the infrared emission (Fig. \ref{fig:circinus_f1}). This interpretation is further discussed in Sect. \ref{sec:disc_circ_bar}.

\subsubsection{CO luminosity and molecular gas mass}\label{sec:circ_lcomgas}

We estimated the CO(3--2) luminosity within a beam positioned at the central position of the galaxy ($r\leq 19.1^{\prime\prime}$), in a 1$^{\prime}$ (1.2 kpc) radius circular aperture (shown in Figure \ref{fig:circinus_totalm0}), and from the area within the dilated mask. The corresponding velocity-integrated fluxes and CO luminosities are listed in Table \ref{tab:fluxes}. For the full galaxy we obtain $L^{\prime}_{\rm CO(3-2)} = (1.5\pm0.4)\times10^{8}$ K km s$^{-1}$ pc$^{2}$.

In order to calculate the molecular gas mass, we must first convert the APEX CO(3--2) luminosity to a corresponding CO(1--0) luminosity, for which we used a line luminosity ratio of $r_{31} = L^{\prime}_{\text{CO(3--2)}}/L^{\prime}_{\text{CO(1--0)}} = 0.31\pm0.11$ computed by \cite{2022ApJ...927..149L}. We note that this is lower than the value of $r_{31} = 0.79$ estimated by \cite{2014A&A...568A.122Z} for the central position ($r\le27^{\prime\prime}$), likely because of AGN excitation, but since our map detects gas well beyond (up to $\sim5^{\prime}$) the \cite{2014A&A...568A.122Z} data and because of beam dilution, we think the \cite{2022ApJ...927..149L} value is more representative of the disk and outskirts of the Circinus galaxy. The sensitivity of our map using this value is $\sim4.9\times10^{6}$ M$_{\odot}$/beam, and the total molecular gas mass of Circinus is $M_{mol} = (2.1\pm1.1)\times10^{9}$ M$_{\odot}$. This is in agreement with the value derived by \cite{2008MNRAS.389...63C} from SEST observations, who derive $M_{mol} =(1.7\pm0.5)\times10^{9}$ M$_{\odot}$.\footnote{\cite{2008MNRAS.389...63C} use an $X_{\mathrm{CO}}$ conversion factor of $(2.3\pm0.3)\times 10^{20}$ cm$^{2}$ (K km s$^{-1})^{-1}$, which corresponds  $\alpha_{\mathrm{CO}} = 5.0\pm0.7$ M$_{\odot}$ (K km s$^{-1}$ pc$^{2})^{-1}$. We have therefore recalculated the molecular gas mass using $\alpha_{\mathrm{CO}} = 4.35\pm1.3$ M$_{\odot}$ (K km s$^{-1}$ pc$^{2})^{-1}$.} From the derived molecular gas mass, we computed standard galaxy parameters for Circinus, including the molecular gas fraction ($\mathit{f}_{mol} = M_{mol}/M_{*}$) and depletion time ($\tau_{dep} = M_{mol}$/SFR), and list these values (for both galaxies) in Table \ref{tab:scaleprops}. In summary, our measurement of $M_{mol}$ confirms the characterization of Circinus as a highly star-forming, gas-rich galaxy.

\begin{figure}
\centering
\includegraphics[width=0.45\textwidth]{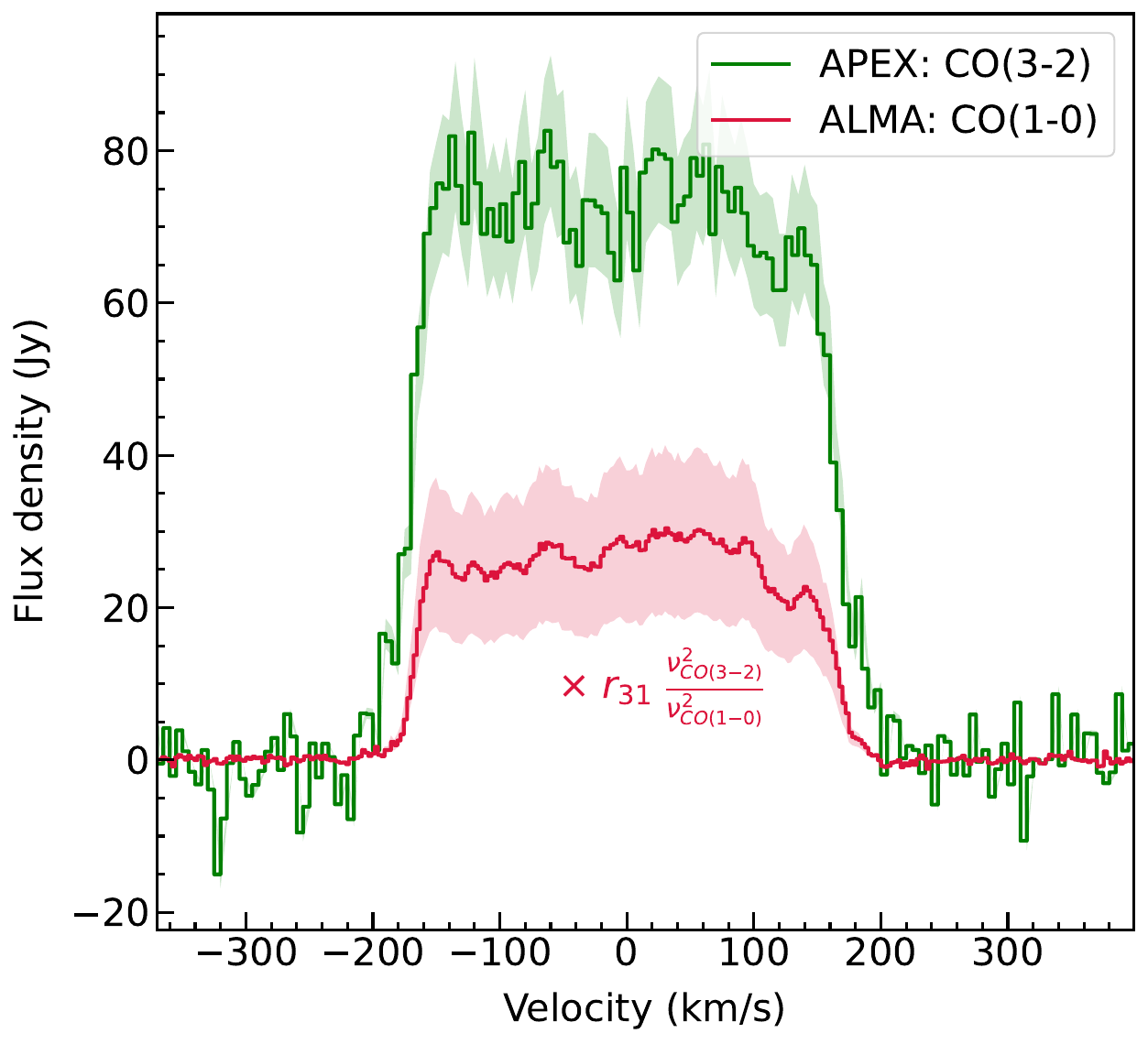}
\caption{Spectra of APEX CO(3--2) and ALMA CO(1--0) (from \cite{2016ApJ...832..142Z}) summed within a $r=1$ kpc ($r=\sim49^{\prime\prime}$) aperture. The ALMA data was rescaled in flux density to CO(3--2) by multiplying a factor of $r_{31} = 0.31\pm0.11$ and the squared ratio of the CO(3--2) ($\nu_{rest} = 345.796$ GHz) and CO(1--0) ($\nu_{rest} = 115.271$ GHz) line frequencies. The light red region shows the 1$\sigma$ error (68\% confidence interval).}
\label{fig:circ_fluxcomp}
\end{figure}

We also compared the fluxes derived from the APEX and ALMA datasets in the 1$^{\prime}$ (1.2 kpc) radius aperture. Since the ALMA data traces a different transition, we rescaled the ALMA CO(1--0) flux density to a CO(3--2) flux density by multiplying by the line luminosity ratio $r_{31}$, and the squared ratio of the line frequencies. The velocity-integrated CO(3--2) fluxes so derived within the 1$^{\prime}$ (1.2 kpc) radius aperture are $S_{\rm CO}\Delta v = (2.5\pm0.3)\times10^{4}$ Jy km s$^{-1}$ for the APEX data and $S_{\rm CO}\Delta v = (9\pm3)\times10^{3}$ Jy km s$^{-1}$ for the ALMA data, which is $\sim35\pm13\%$ ($\sim3$ times lower) than the APEX value. This is further illustrated in the comparison of the APEX and ALMA spectra in Figure \ref{fig:circ_fluxcomp} and in the flux curve-of-growth comparison in Figure \ref{fig:circ_cog}. Despite the uncertainty related to the $r_{31}$ value, this quantifies the extent to which using ALMA 12-m data alone could miss substantial flux even in such inner regions.

\begin{table*}
\centering
\footnotesize
\caption{Integrated flux density and CO luminosity values calculated for Circinus and NGC 1097.}
\label{tab:fluxes}
\begin{tabular}{lcccc}
\hline
\hline\\[-2mm]
Galaxy   & Region           & $S_{\nu} \Delta v$ & $L^{\prime}_{\text{CO}}$ \\ [1mm]
   &            & (Jy km s$^{-1}$) & (K km s$^{-1}$ pc$^{2}$) \\ [1mm]
\hline\\ [-2mm]
Circinus & Central beam ($\theta_{\rm FWHM} = 19.1$\as) & 6950 $\pm$ 790 & $(3.33\pm0.87)\times10^{7}$ \\
CO(3--2) & $r_{\text{aperture}}$ = 1$^{\prime}$ (1.2 kpc) & 25000 $\pm$ 2800 & $(1.19\pm0.31)\times10^{8}$  \\
& Dilated mask & 31600 $\pm$ 3580 & $(1.51\pm0.39)\times10^{8}$ \\
\hline\\ [-2mm]
NGC 1097 & Central beam ($\theta_{\rm FWHM} = 28.7$\as) & 2110 $\pm$ 256 & $(2.37\pm5.82)\times10^{8}$ \\
CO(2--1) & $r_{\text{aperture}}$ = 35$^{\prime\prime}$ (2.3 kpc) & 4100 $\pm$ 500.0 & $(4.59\pm1.13)\times10^{8}$ \\
         & Dilated mask & 6230 $\pm$ 760.0 & $(6.99\pm1.72)\times10^{8}$ \\
         & Tidal features & 128.3 $\pm$ 28.2 & $(1.44\pm0.44)\times10^{7}$ \\ 
\hline
\end{tabular}
\tablefoot{The central beam refers to a beam-sized circular aperture placed at the centers of each galaxy. The dilated mask refers to the masks computed for the galaxies as described in Sect. \S\ref{sec:masking}.}
\end{table*}

\subsubsection{Gas kinematics: Moment maps}\label{sec:circ_kin_moments}

\begin{figure*}
\centering
\includegraphics[width=1\textwidth]{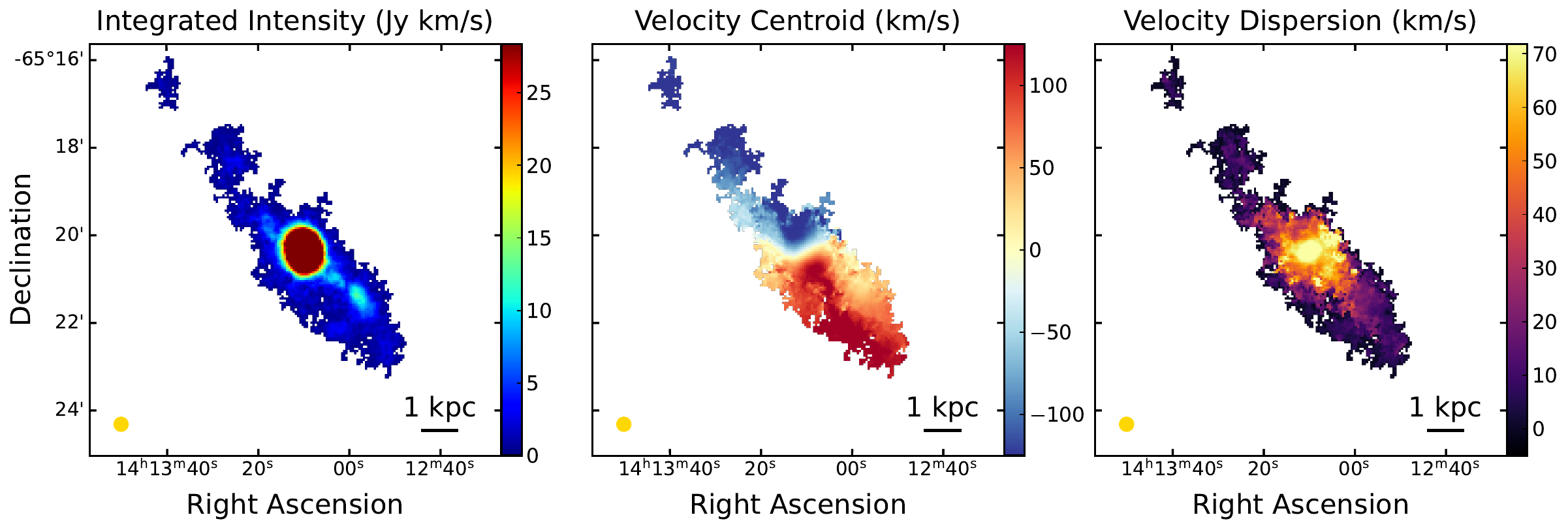}
\caption{Moment maps of Circinus. From left to right: Integrated intensity, velocity centroid, and velocity dispersion (moment 0, 1, 2) maps, obtained by integrating over the data cube after applying the dilated masking procedure, from -370 to 500 km/s. The color scale of the integrated intensity map is kept relatively flat to showcase lower intensity extended structures. The size of the APEX beam is displayed on the bottom left of the panels.}
\label{fig:circ_moments}
\end{figure*}

We plot the integrated intensity, velocity centroid, and velocity dispersion maps in Figure \ref{fig:circ_moments}, computed using the voxels included in the dilated masking technique. The integrated intensity and velocity dispersion maps show a straightforward increase toward the gas-rich nucleus; smaller-scale variations within these regions, such as the detected molecular spiral arms, ring, and outflow \citep{2016ApJ...832..142Z, 2018ApJ...867...48I} are diluted by the APEX beam ($\sim$400 pc at 4.2 Mpc).

The velocity centroid (moment 1) map primarily shows the gas following a rotational pattern. For the fainter gas features at radii larger than 60\as (1.2 kpc), the rotational velocities start decreasing and the position angle of the rotation becomes larger. Such a difference in kinematic behavior of gas at different scales points toward warps in the disk, and is in line with previous observations of the large-scale atomic and molecular gas of Circinus \citep{1998MNRAS.300.1119E, 1999MNRAS.302..649J,2008MNRAS.389...63C}. The dispersion map expectedly shows the largest turbulence ($\sim$85 km/s) in the central region of the galaxy due to the beam size, which cannot resolve the strong velocity gradients in the central region. Overall the dispersion declines smoothly radially from the center, aside from a few sites of increased dispersion at a $\sim$1 kpc radius.

\subsubsection{Gas kinematics: Position-velocity diagrams}\label{sec:circ_kin_pvd}

Position-velocity diagrams provide a simple but rich diagnostic of the kinematic behavior of gas in galaxies. We constructed position-velocity (PV) diagrams using the \texttt{pvextractor} package in \texttt{Python}, which allows for choosing the length, width, and orientation of the slit along which the PV diagram is to be computed. For a simple disk galaxy seen in tracers such as HI, which trace the outermost regions of the disks, the velocity as a function of distance from the center is generally expected to first increase and reach a maximum, after which the velocity generally declines and then remains flat till the outskirts of the galaxy. In position-velocity space for a cut along the major axis, this results in an S-shaped rotation curve. While the molecular gas is expected to trace motions to a shorter extent within the disk, features in the PV diagram of the molecular gas can indicate the presence of noncircular motions, which could arise as a result of molecular outflows \citep{2021A&A...653A.172S}, barred potentials \citep{1999AJ....118..126B}, and molecular rings \citep{2023A&A...675A..88A}.

\begin{figure}
\centering
\includegraphics[width=0.49\textwidth]{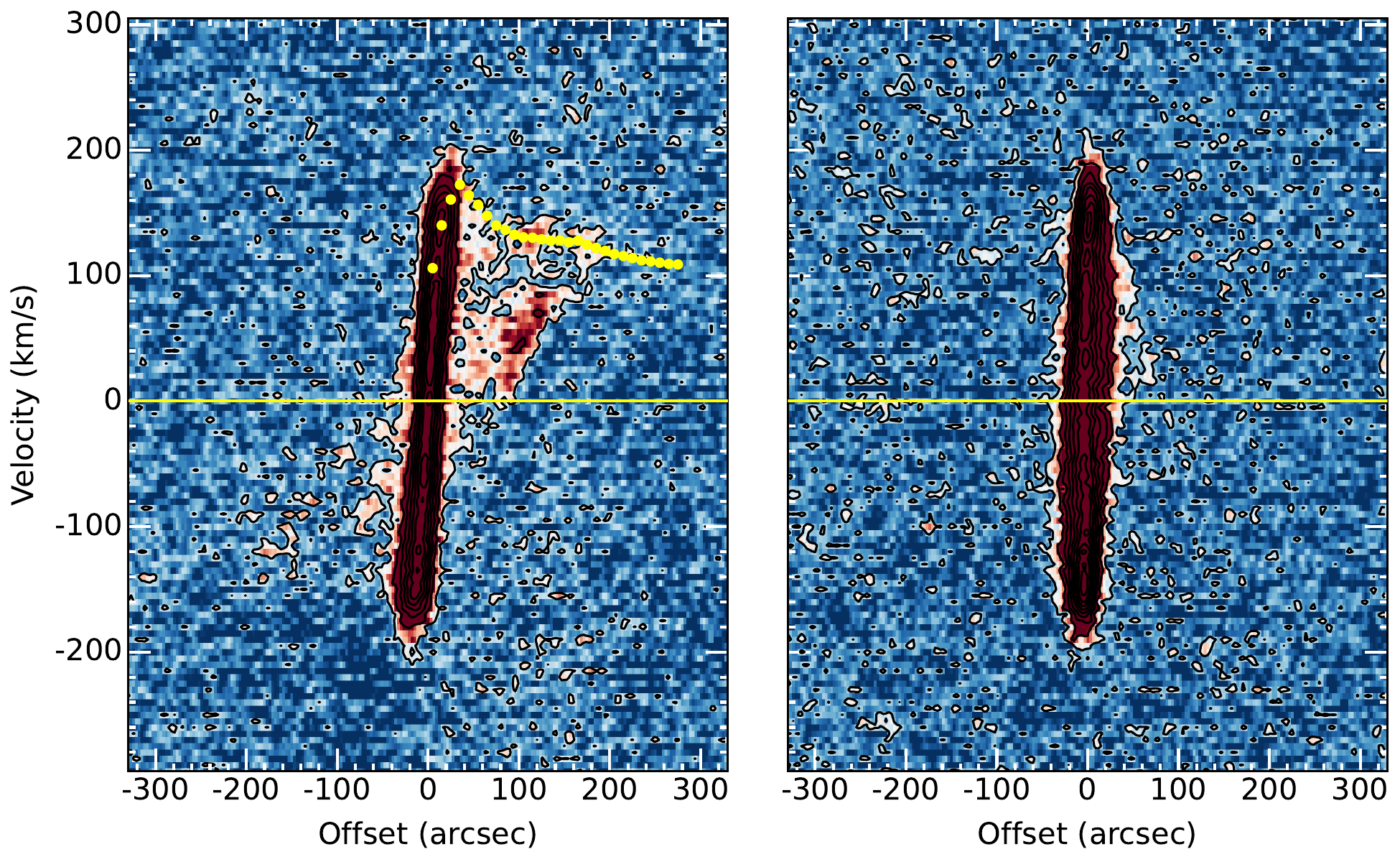}
\caption{Position-velocity diagrams extracted along the major (left) and minor (right) axes of Circinus. Slices were taken along paths of length 11$^{\prime}$, and a width of 2$^{\prime}$ in order to include all the large-scale emission. The major axis slice was aligned at 220$^{\circ}$, and the minor slice at 130$^{\circ}$. The yellow circles mark the projected rotational velocities used in the \bbarolo modeling. Contours in black range from 1.5 to 100 $\times~\sigma$ in steps of 4$\sigma$, where $\sigma$ is 7.5 mJy/pixel. The yellow line marks the systemic velocity of Circinus at 434 km/s.}
\label{fig:circ_pv}
\end{figure}

We extracted PV diagrams along the major and minor axes for Circinus with a length of 11$^{\prime}$ and a width of 2$^{\prime}$ (see Figure \ref{fig:circinus_f1}). The resulting PV diagrams are shown in Figure \ref{fig:circ_pv}. The PV diagram of Circinus along the major axis broadly consists of two components, one being a prominent, steep component which contains the unresolved rotation of the central region of the galaxy (between $\pm60^{\prime\prime}$) terminating at $\sim200$ km s$^{-1}$, and more interestingly, an angled clumpy component  with a projected velocity of $\sim130$ km s$^{-1}$, which corresponds to the large-scale rotation features that are seen in the channels maps and moment maps beyond the center. The minor axis PV diagram is relatively featureless. The presence of the X-shape in the PV diagram in inclined galaxies is suggestive of the presence of resonant orbits followed by the gas, and in turn of noncircular motions \cite{1999AJ....118..126B, 1999ApJ...522..699A}. This is further discussed in Sect. \ref{sec:disc_circ_bar}.

\subsubsection{Gas kinematics: Tilted ring modeling}\label{sec:circ_kin_bbar}

\begin{figure}
\centering
\includegraphics[width=0.5\textwidth]{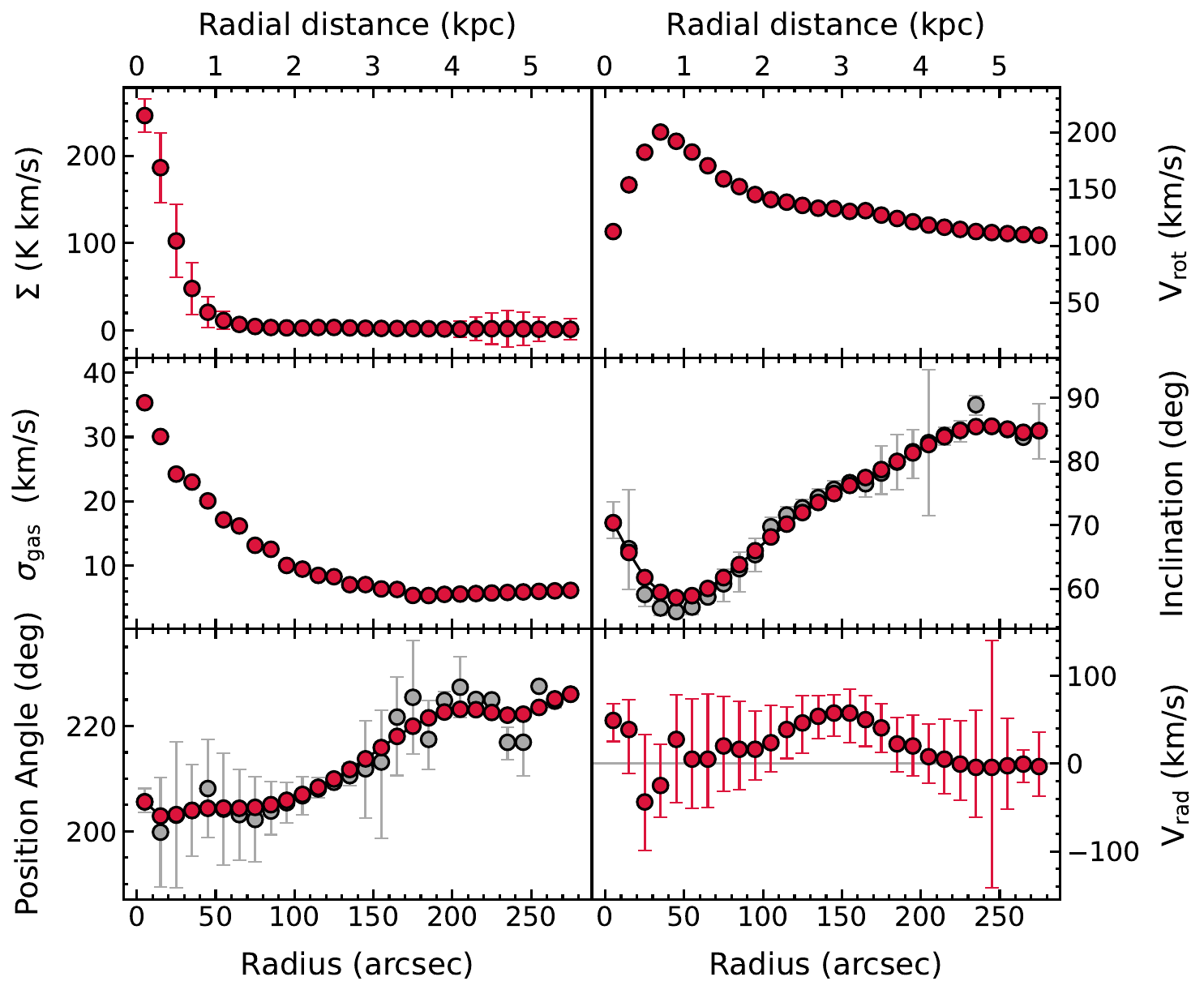}
\caption{Results from the 3D tilted ring modeling of Circinus in $^{\rm 3D}$BAROLO. The markers and error bars show parameters of the fit tilted ring model. For the position angle and inclination, the gray markers with errors show the errors in the first stage of fitting, while the red markers show the parameters after regularization.}
\label{fig:circinus_bb_params}
\end{figure}

The resulting fit parameters for the Circinus data are shown in Figure \ref{fig:circinus_bb_params}, and the resulting moment maps and residuals for both models are shown in Figures \ref{fig:circinus_bb_novrad} and \ref{fig:circinus_bb_vrad}. Both rotation and dispersion curves were adjusted as described in \ref{sec:method_barolo}. The rotation curve was set to match the PV diagram along the major axis (see Figure \ref{fig:circ_pv}), increasing from 100 km/s to 200 km/s, and then flattening out to $\sim110$ km/s. These velocities are also consistent with the rotation curve found by \cite{2008MNRAS.389...63C}. The velocity dispersion curve was set to decline smoothly from about 35 km/s in the center to 7 km/s at the outskirts of the galaxy. 

The inclination and position angle, which were regularized to follow bezier curves in a two-stage fitting process, show large variations across the disk of Circinus. The inclination angle follows a decline from the center till 50\as after which it increases toward the outskirts of the disk, while the position angle follows a gradual increase from the center. The total variation of about 24$^{\circ}$ in the position angle and 27$^{\circ}$ in the inclination angle, and similar total variations seen over the entire 70 kpc HI disk of Circinus, underscores the warped nature of Circinus \citep{2008MNRAS.389...63C}.

After fixing all these parameters, we fit a model which includes a radial velocity component, the result of which is shown in the last panel of Figure \ref{fig:circinus_bb_params}. While we obtain positive radial velocities across the galaxy (going up to roughly 40-50 km/s between 115\as to 175$^{\prime\prime}$), the error bars are of the same order. Hence no significant radial velocities are detected in this data. We further discuss the implications of the radial fit for Circinus in Sect. \ref{sec:disc_tilted_modeling}.

\begin{table}
\centering
\footnotesize
\caption{Physical properties of Circinus and NGC 1097.}
\label{tab:scaleprops}
\begin{tabular}{lccc}
\hline
\hline \\ [-2mm]
        Property & Circinus & NGC 1097 \\ [1mm]
\hline \\ [-1mm]
$\log{(M_{mol})}/[M_{\odot}]$ & 9.33 & 9.67 \\[1mm]
$\log{(M_{*})}/[M_{\odot}]$ & 10.98 & 10.76 \\[1mm]
$\log{(\text{SFR})}/[M_{\odot}~yr^{-1}]$ & 0.63 & 0.68 \\[1mm]
$\log{(\mathit{f}_{mol})}$ & -1.65 & -1.09  \\[1mm]
$\log{(\tau_{dep})}$      & 8.69 & 8.99 \\[1mm]
\hline
\end{tabular}
\tablefoot{The molecular mass gas fraction listed here is defined as the molecular gas mass ($M_{mol}$) divided by the stellar mass ($M_{*}$). References for the $M_{*}$ and SFR values are listed in Table \ref{tab:galprops}.}
\end{table}

\subsection{NGC 1097}\label{sec:ngc1097_results}
\subsubsection{Gas distribution}\label{sec:ngc1097_gasdist}
\begin{figure}
\centering
\includegraphics[width=0.49\textwidth]{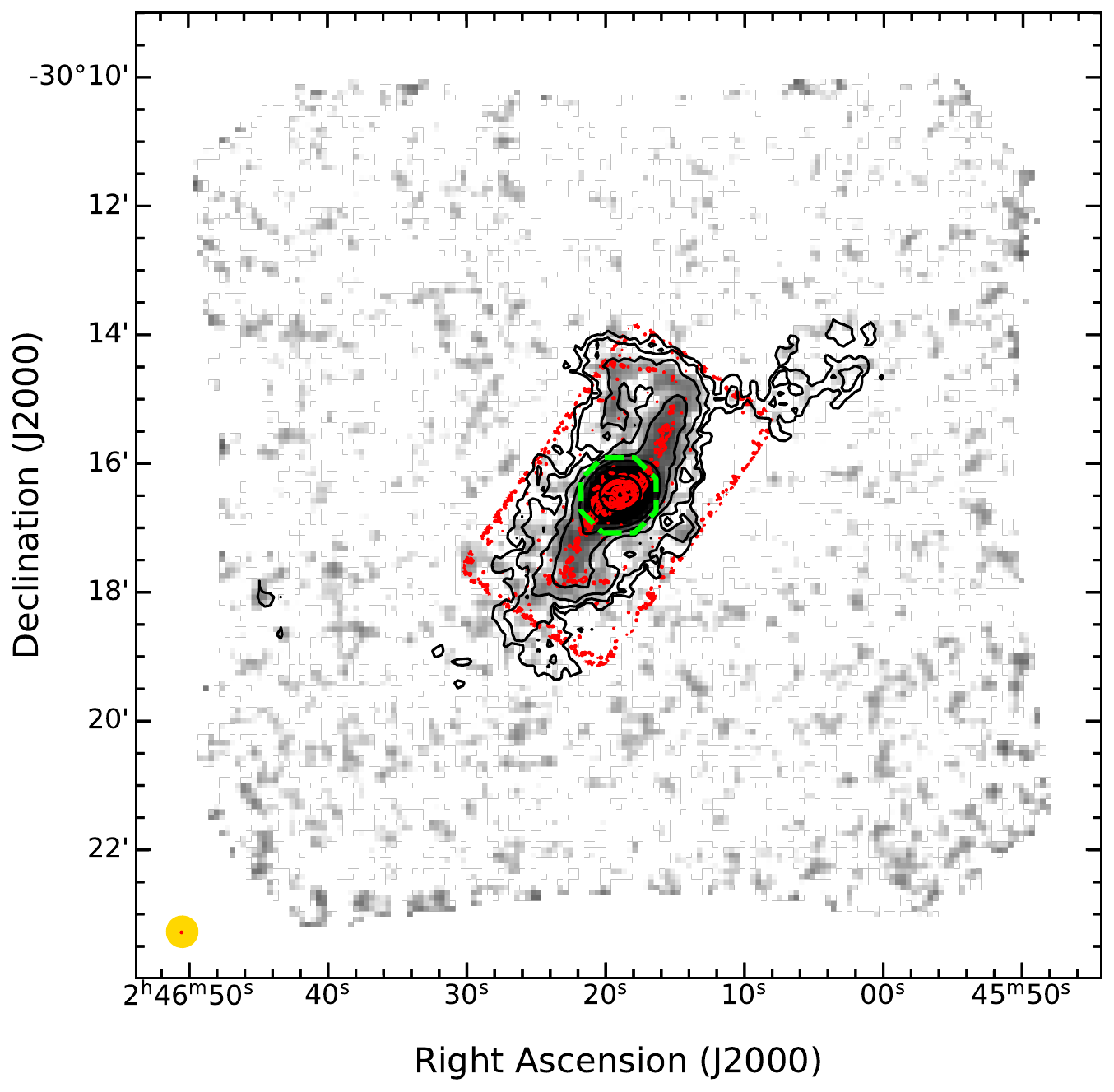}
\caption{CO(2--1) integrated intensity map computed from the unmasked data from $-$285 to 410 km/s for NGC 1097. The APEX contours are plotted from the masked integrated intensity map in 8 log-spaced levels from 0.15 Jy km/s to 75.6 Jy km/s. The red contours show the integrated intensity map of the ALMA CO(2--1) data from \cite{2021ApJS..257...43L}. The 4 contours are plotted on log-spaced intervals, starting at 1.9 Jy/beam km/s to 24.7 Jy/beam km/s. The dashed green circle marks the 35\as (2.3 kpc) radius aperture around the starburst ring. The beam size of the APEX and ALMA observations is shown in the bottom left.}
\label{fig:ngc1097_totalm0}
\end{figure}

Figure \ref{fig:ngc1097_f1} shows (in contours) the integrated intensity map of NGC 1097. The center, which comprises the bright starburst ring, is unresolved and we do not see any substructure within this region. The lower intensity emission beyond the central few kpc clearly shows the bar that feeds the starbursting nuclear ring, as well as its termination in spiral arms. Toward the northwest, there is a prominent feature azimuthally offset from the direction of the spiral arm coinciding with the disrupted spiral arm seen in the \textit{Spitzer} map, which we identify as a tidal feature. This is also the furthest detection of gas in the map, at about 4.5$^{\prime}$ ($\sim$18 kpc) from the nucleus. Toward the southeast, the furthest detection is at roughly 3$^{\prime}$ ($\sim$14 kpc), where the bar begins to wind into the outer spiral arm. We can also clearly identify the CO emission associated with the tidal structure in the northwest in the integrated intensity map.

In Figure \ref{fig:ngc1097_totalm0}, we show a comparison between the unmasked integrated intensity map of the APEX data, overlaid with contours from the masked data (identical to those in Figure \ref{fig:ngc1097_f1}), and the PHANGS-ALMA data (in red contours), which in this case traces the same CO transition as the APEX data. Unlike in the case of Circinus, the PHANGS-ALMA map seems to cover most of the gas emission from the galaxy; however, the tidal feature still lies outside the area covered by ALMA toward the northwest.

\subsubsection{CO luminosity and molecular gas mass}\label{sec:ngc1097_lcomgas}
We calculated the CO(2--1) luminosity in NGC 1097 within a beam at the central position ($r\leq28.7^{\prime\prime}$), a 35\as (2.3 kpc) radius circular aperture enclosing the bright starburst ring in the center, and from the total area defined within the dilated mask (listed in Table \ref{tab:fluxes}). For the latter, we derive a CO(2--1) line luminosity of $L^{\prime}_{CO(2-1)}= 6.99\pm1.72 \times 10^{8}$ K km s$^{-1}$ pc$^{2}$. This is $\sim$82$\pm20\%$ of the ALMA derived CO luminosity from \cite{2021ApJS..257...43L}.

\begin{figure}
\centering
\includegraphics[width=0.49\textwidth]{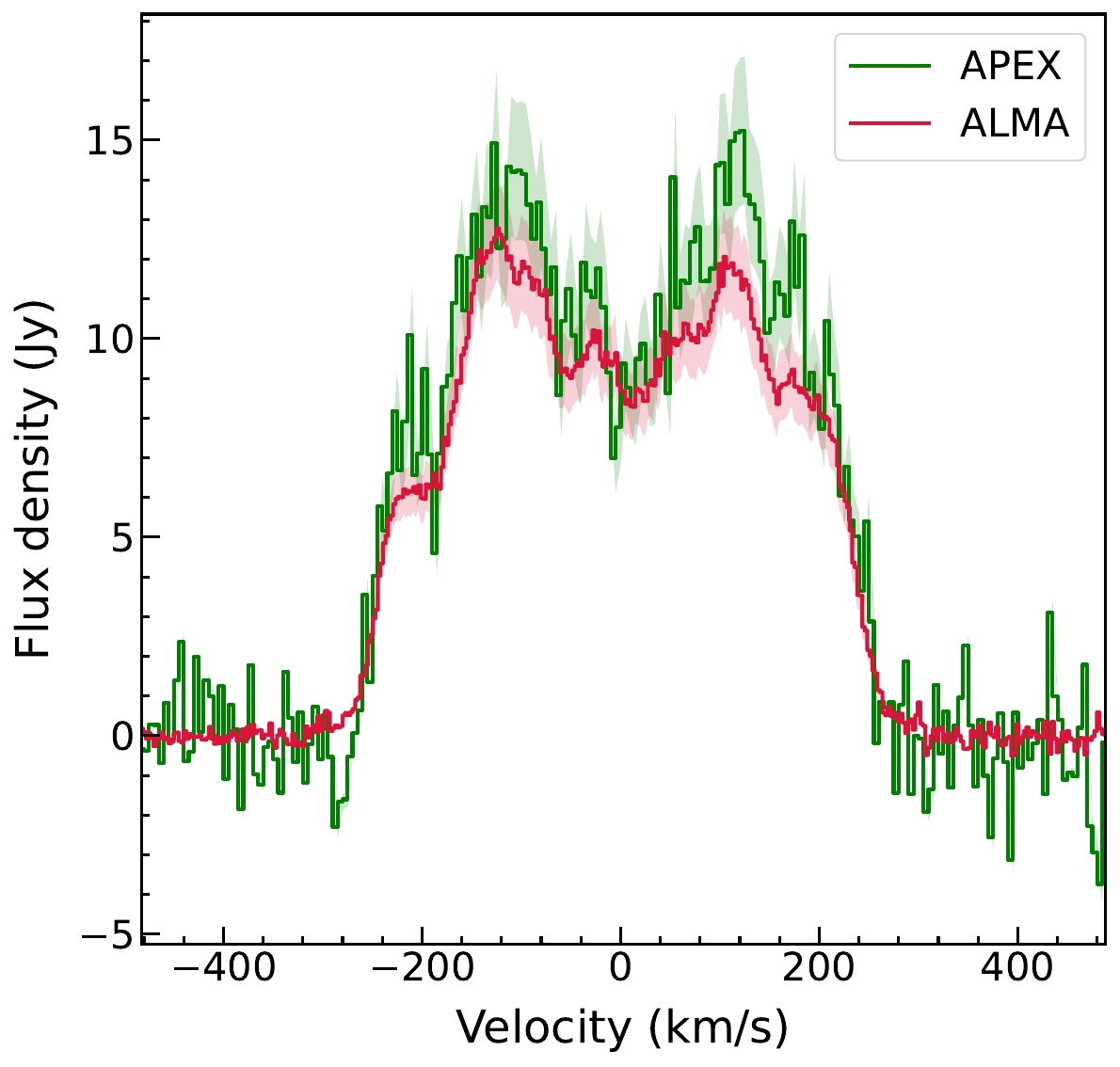}
\caption{Spectra of APEX CO(2--1) and ALMA CO(1--0) (from \cite{2021ApJS..257...43L}) integrated within a 35\as (2.3 kpc) radius aperture. The light red region showcases a 1$\sigma$ error in the flux density, which mostly originates from the 10\% calibration uncertainty of ALMA Band 6 data.}
\label{fig:ngc_fluxcomp}
\end{figure}

We computed the total molecular gas mass of NGC~1097, as well as the mass within the region of the starburst ring. We assumed a $L^{\prime}_{CO(2-1)}/L^{\prime}_{CO(1-0)}$ ratio of $r_{21}=0.65\pm0.18$ representative of local star-forming galaxies, from \citet{2022ApJ...927..149L}. However, we note that the $r_{21}$ parameter for NGC 1097 has been found to be roughly unity in the bar region, and larger than unity in the starburst ring \citep{2008ApJ...683...70H}, therefore the mass we derive under the assumption of a lower $r_{21}$ ratio could be overestimated, especially in the central region. Assuming a galactic $\alpha_{\text{CO}}$ conversion factor, the sensitivity of the map is $\sim1.8\times10^{7}$ M$_{\odot}$/beam, and the total molecular gas mass in the dilated mask region is $M_{mol} = (4.7\pm2.2)\times10^{9}$ M$_{\odot}$. The mass within the central region including the starburst ring is $(3.1\pm1.5)\times10^{9}$ M$_{\odot}$. \cite{2001PhDT.......237C} find a lower limit on the total molecular gas mass using the CO(1--0) line at $M_{mol}\approx (6\pm2)\times10^{9}$ M$_{\odot}$\footnote{The $M_{mol}$ value has been corrected for the different distance used by \cite{2001PhDT.......237C}, who use $D_{L}$ = 17 Mpc for NGC~1097.}, which is consistent within the uncertainties with our measurement. Translating the \cite{2021ApJS..257...43L} $L^{\prime}_{CO}$ measurement using the galactic $\alpha_{CO}$ factor and $r_{21} = 0.65$ gives $M_{mol} = (5.7 \pm 1.8) \times 10^{9}$ M$_{\odot}$, which is also consistent with our $M_{mol}$ measurement.

\subsubsection{Gas kinematics: Moment maps}\label{sec:ngc1097_kin_moments}

\begin{figure*}
\centering
\includegraphics[width=1\textwidth]{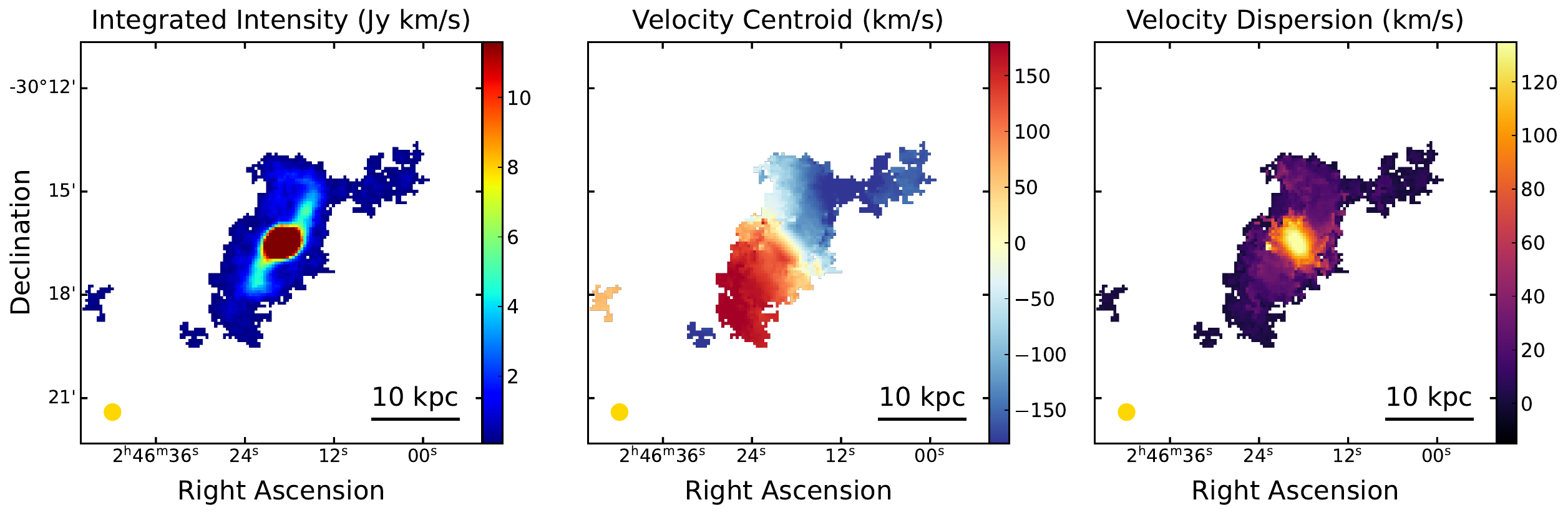}
\caption{Moment maps of NGC 1097. From left to right: Integrated intensity, velocity centroid, and velocity dispersion (moment 0, 1, 2) maps, integrated over the full range of the data cube after applying the dilated masking procedure, from -485 to 500 km/s. The size of the APEX beam is shown on the bottom left of the images.}
\label{fig:ngc_moments}
\end{figure*}

The integrated intensity, velocity centroid, and velocity dispersion maps for NGC 1097 are presented in Figure \ref{fig:ngc_moments}. The integrated intensity map is discussed in Sect. \ref{sec:ngc1097_gasdist}. The velocity centroid map clearly shows a rotational pattern on all scales, and the position angle remains constant unlike the case of Circinus. The two small clumps seen slightly toward the southeast and toward the eastern edge of the map are clearly kinematically incompatible with the nearby emission, and thus we consider these to be artifact emission. The tidal feature seen in the northwest in the receding gas on the other hand is coherent in velocity with the rest of the galaxy.

The velocity dispersion map reflects the turbulence associated with the central starburst ring, with peak velocity dispersion values close to 150 km/s, which is due to the large unresolved velocity gradients in the center. Beyond this central region, the gas exhibits much lower $\sigma_v$ closer to 20-30 km/s, and this gas also traces the bar and their winding close to the outer spiral arms.

\subsubsection{Gas kinematics: Position-velocity diagrams}\label{sec:ngc1097_kin_pvd}

\begin{figure}
\centering
\includegraphics[width=0.49\textwidth]{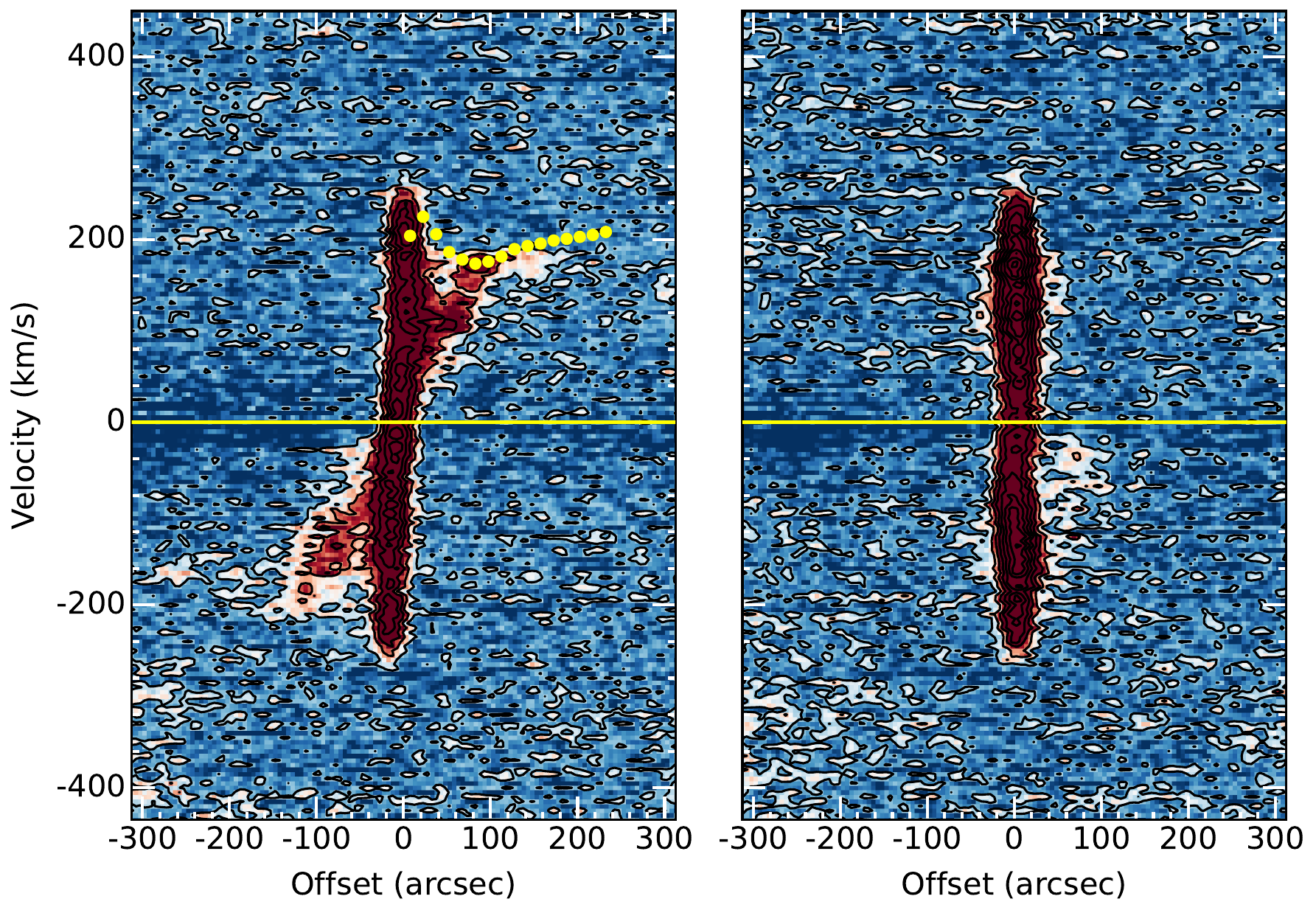}
\caption{Position-velocity diagrams extracted along the major (left) and minor (right) axes of NGC 1097. Slices were taken along paths of length 11$^{\prime}$, and a width of 4$^{\prime}$ in order to include all the large-scale emission. The major axis slice was aligned at 137$^{\circ}$, and the minor slice at 47$^{\circ}$. The yellow circles mark the projected rotational velocities used in the \bbarolo modeling. Contours in black range from 1 to 100 $\times~\sigma$ in steps of 3$\sigma$, where $\sigma$ is 1.8 mJy/pixel. The yellow line marks the systemic velocity of NGC 1097 at 1258 km/s.}
\label{fig:ngc1097_pv}
\end{figure}

We plot PV diagrams along the major and minor axes of NGC 1097, with position angles of 137$^{\circ}$ and 47$^{\circ}$, respectively, and with widths of 4$^{\prime}$ to cover all the emission in the mask (see Figure \ref{fig:ngc1097_f1}). The resulting PV diagrams are shown in Figure \ref{fig:ngc1097_pv}. In this case, similar to Circinus, we see a clearer X-shape in the major axis PV diagram, which is expected given the prominent bar of NGC 1097. The minor axis PV diagram is relatively featureless, aside from some low brightness features at $v=-40$ km/s, which appears to be gas in the region between the bar and the inner ring (also seen close to the central emission in the $v=-60, -10$ km/s channel maps in Figure \ref{fig:ngc_chanmap}).

\subsubsection{Gas kinematics: Tilted ring modeling}\label{sec:ngc_kin_bbar}

\begin{figure}
\centering
\includegraphics[width=0.5\textwidth]{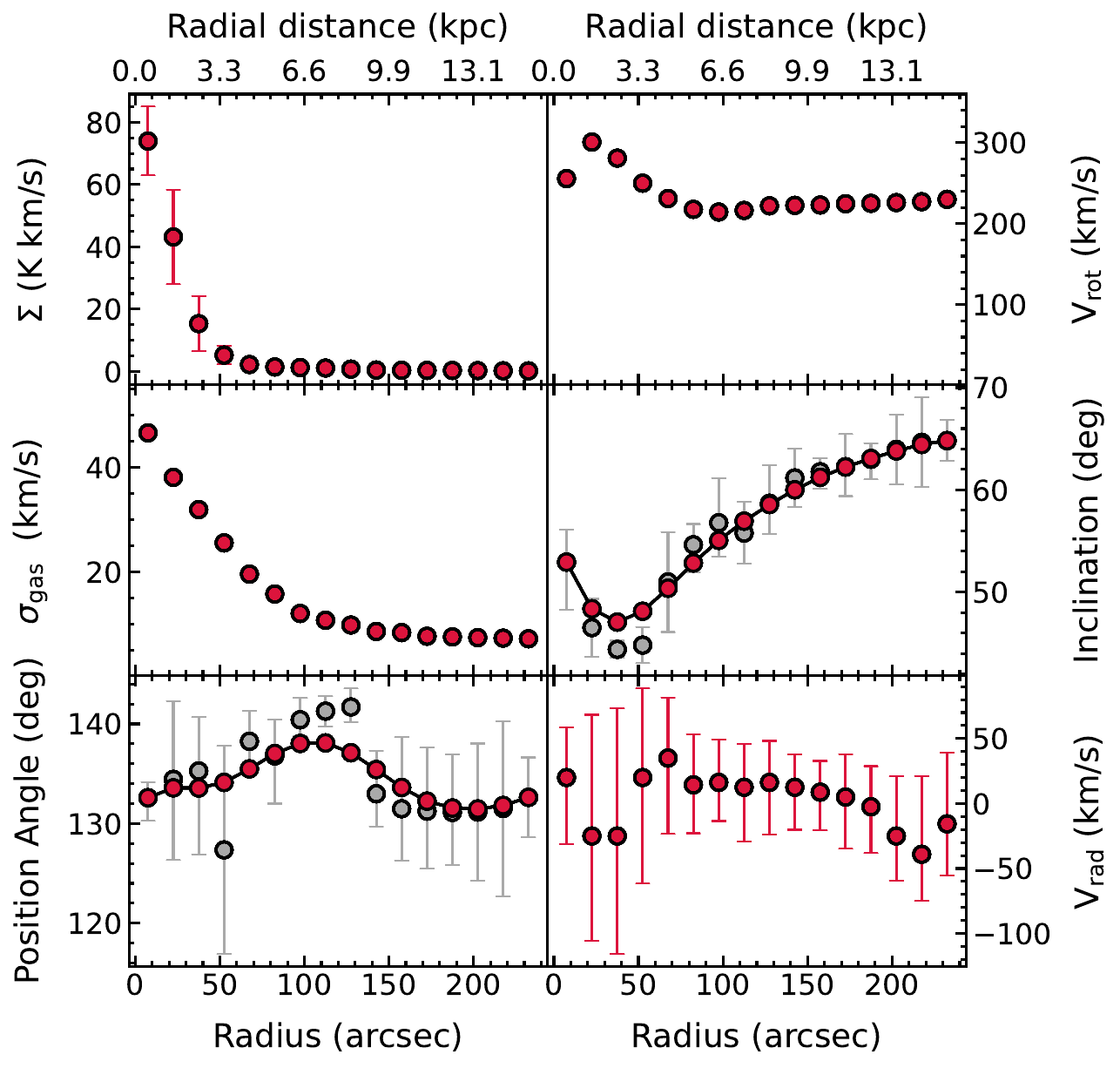}
\caption{Results from the 3D tilted ring modeling of NGC~1097 in $^{\rm 3D}$BAROLO. The markers and error bars show parameters of the fit tilted ring model. For the position angle and inclination, the gray markers with errors show the errors in the first stage of fitting, while the red markers show the parameters after regularization.}
\label{fig:ngc1097_bb_params}
\end{figure}

The resulting model parameters found for NGC~1097 are shown in Figure \ref{fig:ngc1097_bb_params}, and model maps with residuals are shown in Figures \ref{fig:ngc1097_bb_novrad} and \ref{fig:ngc1097_bb_vrad}. The adjusted rotation curve obtained for NGC~1097 is similar to the one for Circinus. In the inner 40$^{\prime\prime}$ the de-projected velocity peaks at about 300 km/s, after which the velocity declines and stays flat at about 220 km/s in the outer regions. Since the velocity dispersion like Circinus artificially jumps to very high values, the profile for the dispersion was chosen such that it declines from about 45 km/s to 5 km/s from the center to the outskirts.

Similar to Circinus, in NGC 1097 the inclination first declines till $\sim$40\as and then increases substantially toward the outskirts of the disk, while the position angle shows a smaller increase and decline. The total variation in the inclination angle and position angle are about 15$^{\circ}$ and 8$^{\circ}$, respectively, thus exhibiting a smaller warp than Circinus. As for Circinus, the derived radial velocities have high uncertainties, and are therefore not significant. We further discuss the implications of the kinematic modeling in NGC~1097 in Sect. \ref{sec:disc_tilted_modeling}.

\section{Discussion}\label{sec:discuss}

\subsection{Large-scale features in Circinus: molecular bar or spiral-arm signatures}\label{sec:disc_circ_bar}

The disturbed structures in the southwest of Circinus have been discussed by \cite{1998MNRAS.300.1119E} and \cite{2008MNRAS.389...63C}, who note that these structures show abrupt position angle changes in both atomic and molecular gas at $>10$ kpc and $>1$ kpc, respectively. They suggest these arise from sudden changes in the path of the molecular gas traversing the dust lane of a barred potential \citep{1999ApJ...526...97R, 1999ApJ...525..691S, 2001ApJ...547..792D}. \cite{1999MNRAS.302..649J} also find evidence in the HI gas dynamics within $r<250$\as that suggests the presence of a 10 kpc bar at a position angle of 225$^{\circ}$. While the presence of a molecular bar and noncircular kinematics of the molecular gas has been discussed in the $\sim400$ pc inner molecular ring \citep{2000ApJ...531..219M, 2018ApJ...867...48I}, a similar discussion about the presence of such a molecular structure in the extended emission is absent. The larger FoV and sensitivity of our APEX map includes additional emission in the southwest than that seen by \cite{2008MNRAS.389...63C}, and enables us to discuss the southwestern emission in this context.

The Spitzer 8 $\mu$m dust emission seen in Figure \ref{fig:circinus_f1} shows a curved dust lane in the region associated with the extended CO emission seen in the APEX data. Purely morphologically, the curved dust lanes can suggest the presence of a spiral arm or a bar feature. This dust lane harbors bright clumps that also have elevated levels of atomic hydrogen, CO(1--0), and CO(2--1) emission (\cite{2012MNRAS.425.1934F} see C-1, C-2, in their Figure 12). The atomic gas distribution seen in this region was classified by \cite{1999MNRAS.302..649J} as being part of an atomic gas bar with a radius of $\sim$5 kpc. In Figure \ref{fig:circinus_f1}, the molecular gas emission appears to follow the curved dust lanes, which suggests that the molecular gas gets compressed as it enters the density wave, and therefore leads to local star formation. While star formation in bars is thought to be suppressed due to the high shear experienced by gas in barred orbits, there is evidence of star formation being induced in weaker bars due to the shocking of gas and weaker shear forces \citep{2002ApJ...570L..55J, 2020A&A...644A..38D, 2021A&A...654A.135D}. The latter case is what we appear to be seeing in the case of Circinus, since the curvature of dust lanes in bars has also been found to anti-correlate with the strength of shear forces in bars \citep{2009ApJ...706L.256C}. This suggests that the bar in Circinus would be weak given the curved dust lanes seen in the dust continuum.

Also, an ``X-shaped'' position-velocity diagram has previously been suggested as an indicator of the presence of a bar \citep{1999ApJ...522..686B, 1999ApJ...522..699A}, in ionized \citep{1999AJ....118..126B, 1999A&A...345L..47M} and molecular gas \citep{1992ApJ...399...94W, 2013MNRAS.432.1796A, 2018MNRAS.476..122V, 2023A&A...675A..88A}, due to the difference in orbits followed in a barred potential. Specifically, these correspond to two different families of periodic orbits, namely \textit{x$_{1}$} and \textit{x$_{2}$} orbits. The \textit{x$_{1}$} orbits are elongated along and mainly form the bar structure, have longer rotation periods, are mostly elliptically shaped, while the \textit{x$_{2}$} family of orbits are located closer to the center (corresponding to the Inner Lindblad Resonance), are elongated slightly perpendicular to the bar, and exhibit faster rotation. The consequence of this in a PV diagram for highly inclined galaxies is that the steeper and hence faster rotating component corresponds to rotation closer to the center of the galaxy, while gas orbiting in the bar develops a shallower slowly rotating component.

As discussed in Sect. \ref{sec:circ_kin_pvd}, our APEX data confirms the ``X-shape'' feature in the PV diagram of Circinus, which have previously been observed by \cite{1998MNRAS.300.1119E} and \cite{2008MNRAS.389...63C}. The observed steeper component is understood to correspond to a molecular ring in the center of the galaxy, which is studied in detail in \cite{2018ApJ...867...48I, 2023Sci...382..554I}. The presence of such a signature in the PV diagram of Circinus, the correspondence of the shallower component with the extended emission, and the star formation seen in this region suggest that the extended molecular gas emission seen by APEX arises from a weak molecular bar, rather than a loosely wound spiral arm.

\subsection{Implications for the molecular outflow in Circinus}\label{sec:circ_outflow?}

\cite{2016ApJ...832..142Z} discuss the presence of a cold molecular phase of the outflow of Circinus using ALMA CO(1--0) data. We checked the APEX data for signatures of an outflow in integrated spectra positioned at their identified candidate regions, but do not see signs of blue-shifted outflow emission. The velocity dispersion in the inner $1^{\prime}$ seen by ALMA is also found to be higher ($\sim30-50$ km s$^{-1}$, compared to $\le$ 10 km s$^{-1}$ for disk gas) in one of the candidate regions; however, while the APEX dispersion map does show enhanced values of $\sim 70-80$ km s$^{-1}$ in the center and toward the northwest of the disk, these magnitudes can be attributed to the APEX beam not resolving large velocity gradients, and the enhancement toward the northwest could also be due to the masking procedure picking low S/N points at the edges of the mask.

Therefore the resolution and sensitivity of the APEX data are insufficient to see the potential outflow features discussed by \cite{2016ApJ...832..142Z}, who also note that the existence of the molecular phase of the outflow is ambiguous even with the ALMA data, due to the outflow and underlying disk having similar projected velocities.

\subsection{Tidal structures in NGC 1097}\label{sec:ngc1097_tidals}
\begin{figure}
\centering
\includegraphics[width=0.49\textwidth]{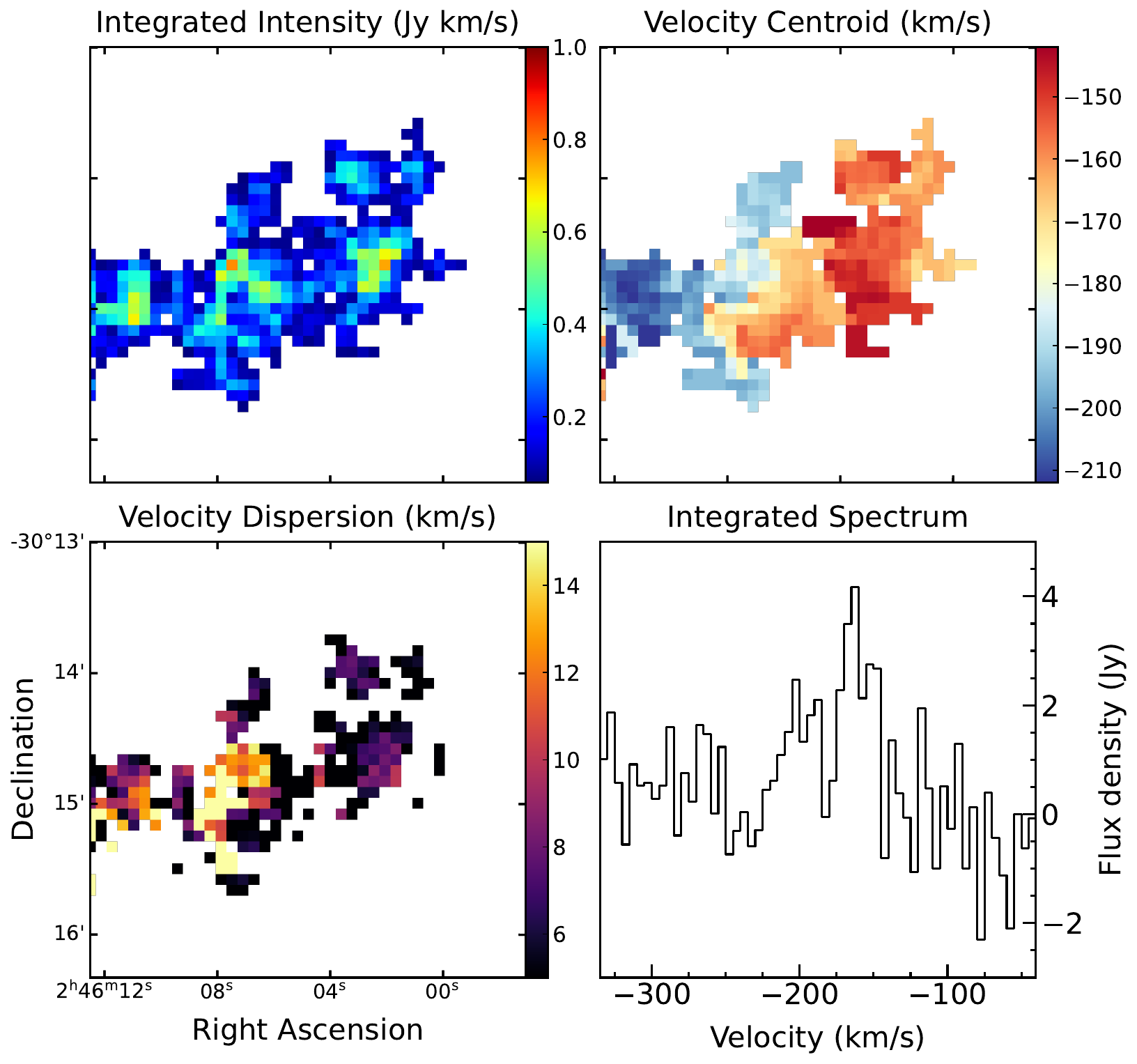}
\caption{Moment maps (first three panels) and integrated spectrum (bottom right) of tidal features in NGC 1097. The color map of the velocity centroid map is rescaled to the median velocity of the feature ($v = -177$ km/s) to reflect the gradient in velocity. The length of the feature is approximately 10 kpc.}
\label{fig:ngc1097_tidals}
\end{figure}

In the integrated intensity and channel maps of NGC 1097, (Fig. \ref{fig:ngc1097_f1}, Fig. \ref{fig:ngc_chanmap}), we detect molecular gas toward the northwest of the galaxy. Figure \ref{fig:ngc1097_tidals} shows the moment maps associated with this feature, as well as a spectrum extracted from the region shown in the panels. Despite the lower surface brightness of this structure, we believe the emission arising is real based on the coincidence with peaks in the Spitzer continuum map, the HI maps \citep{1989ApJ...342...39O, 2001PhDT.......237C, 2003ApJ...585..281H}, in optical images, as well as a clear spectral detection of the feature in the bottom-right panel of Fig. \ref{fig:ngc1097_tidals}. It is noteworthy that although the two peaks in the integrated intensity CO(3--2) map are roughly similar, the HI emission is stronger toward the western peak, while the optical and dust continuum peaks are stronger toward the eastern peak. The molecular gas mass entrained in these tidal features is $(10\pm4)\times10^{7}$ M$_{\odot}$, which constitutes $\sim 2$\% of the total molecular mass measured in the galaxy. The location of this feature coincides with dusty regions seen in the Spitzer 8 $\mu$m continuum and the disturbed spiral arm of NGC 1097 which is being disrupted by the presence of the interacting galaxy NGC 1097A (yellow circle in Fig. \ref{fig:ngc1097_f1}). However, we do not see much molecular gas detected in other positions in the outer spiral arms of the galaxy. This points toward a scenario where the features might be the result of an interaction with NGC 1097A, which has triggered the formation of molecular gas and resulted the formation of stars, or that the gas has been stripped away from the host galaxy.

We can get some insight into the nature of star formation in this region by comparing the molecular gas depletion time ($\tau_{dep} = M_{mol}/{\rm SFR}$) of the tidal tail compared to the full galaxy. The SFR can be calculated using the Wide-field Infrared Survey Explorer (WISE) band 4 map at 22 $\mu$m of NGC~1097 using the conversion from $\nu L_{\nu}$ to SFR used in \cite{2021ApJS..257...43L} for WISE band 4 luminosities, i.e., by multiplying a factor $C = 10^{-42.63} M_{\odot} \ {\rm yr}^{-1} ({\rm ergs \ s^{-1})}^{-1}$. Using this, we obtained SFRs of $4.4\pm1.4$ $M_{\odot} \ {\rm yr^{-1}}$ for the total galaxy (within the dilated mask), and $0.06\pm0.03$ $M_{\odot} \ {\rm yr^{-1}}$ for the tidal tail (in the region shown in Figure \ref{fig:ngc1097_tidals}). Dividing the corresponding gas masses by these values, we obtained depletion times of $0.8\pm0.4 \rm \ Gyr$ and $1.6\pm0.8 \rm \ Gyr$ for the whole galaxy and tidal tail, respectively. This suggests that the tidal tails are less efficient at forming stars than on average in the galaxy. However, we caution that there are many assumptions involved in this calculation, since linear conversion factor \textit{C} is typically calibrated for SFR and $\nu L_{\nu}$ values of full galaxies, and is therefore less applicable for diffuse regions such as tidal tails.

Looking at the integrated intensity map in Fig. \ref{fig:ngc1097_tidals}, these tidal features show two clear intensity peaks, which also coincide with the peaks in these locations seen in the \textit{Spitzer} 8 $\mu$m continuum map. The velocity map shows a gradient in the direction away from NGC 1097, so that even though the bulk of the gas in the tidal tail is rotating with the rest of the galaxy (see Fig. \ref{fig:ngc_moments}), there are internal gas motions associated with the molecular clouds in these regions. The gradient in the first-moment map is reflected in the double-peaked nature of the integrated spectrum seen in the region. The dispersion map exhibits a strong peak toward the eastern star-forming region which indicates higher turbulence in the region which might be the result of feedback or instabilities within the galaxy, or due to the tidal interaction. We looked for potential shock signatures using the SiO(5--4) line \citep{1997A&A...321..293S}, which is within the coverage of the APEX nFLASH bands, but we do not detect it. A molecular gas component in tidal tails has been found in nearby merging galaxies (see Section 3.1 of \cite{2022ApJ...936L..11S} and references therein, \cite{2013LNP...861..327D}). For NGC~1097 the small fraction (2\%) of molecular gas mass contained in the tidal tail means the overall star formation rate is largely unaffected. Nevertheless the tidal tail might offer in a unique environment in which to study the conditions for star formation.

A notable feature seen in optical images of NGC 1097 is the presence of four optical ``filaments'' or ``jets'' \citep{1975MNRAS.173P..51W, 1976ApJ...207L.147A, 1978ApJ...222L..99L} seemingly originating from the nucleus toward the northeast and southwest, which are thought to be the result of activity from the AGN, due to in situ star formation, or results of an interaction with NGC 1097A. \cite{2003ApJ...585..281H} use deep HI observations with the VLA and find the jets to be HI poor and rule out these scenarios and instead suggest a past minor-merger scenario with a dwarf galaxy to reproduce the jets. In our maps, which cover the area including the optical jets, we do not see any CO(2--1) emission associated with those positions.

\subsection{Results of tilted ring modeling and applicability to APEX data}\label{sec:disc_tilted_modeling}

Tilted ring models have been extensively used in the past to model gas kinematics, and to in turn trace dynamical processes in disks galaxies \citep{1971A&A....13...99R, 1973A&A....29..447A, 1974ApJ...193..309R}. Such models have typically been applied to HI observations, which can reach large radii up to twice the optical diameter of the galaxy. More recently, the high resolution and sensitivity offered by telescopes such as ALMA have enabled the application of such modeling to molecular gas observations to disentangle the kinematic signatures of noncircular motions from the disk rotation \citep{2021ApJ...923..220D, 2023A&A...675A..88A, 2024MNRAS.530..446H}. The presence of radial flows is found to be correlated with the presence of bar-like structures and tidal interactions. However these flows are transient and occur on shorter timescales than the dynamical timescales of bars and spiral arms, so the presence of a bar does not necessarily imply the presence of molecular gas inflows \citep{2024MNRAS.528.6768H, 2023ApJ...944..143W}.

Detecting the presence of radial flows in a galaxy such as Circinus is additionally challenging given the high inclination of the galaxy, and that the bar along which one would expect radial flows to be prominent is aligned along the major axis, where the detection of radial flows is difficult due to them being orthogonal to the line of sight. The high inclination in turn reduces information along the minor axis where disentangling the effects of warps in the position angle and inclination with radial flows is easier \citep{2023ApJ...944..143W, 2021ApJ...923..220D}. The addition of the radial velocity component only leads to a small improvement in the residuals in the regions corresponding to the bar seen in the southwest. A promising aspect of the derived radial velocities (final panel, Figure \ref{fig:circinus_bb_params}) is that they are highest at $\sim$115$^{\prime\prime}$-175$^{\prime\prime}$ where the relative error in the radial velocity is also lowest.

In the case of NGC~1097, which is a prototypical barred galaxy, we expect to see a more substantial impact of adding a radial velocity component. Comparing the residuals of the two models, however, such an improvement is not very apparent. This could be due to the large beam size of the data (33$^{\prime\prime}$ about 8$\times$4 beams along the major and minor axes, respectively), and the large errors associated with the fit geometry, in particular the position angle. The result of our application of \bbarolo is that the sensitivity and resolution at these scales of the APEX data is insufficient for pinpointing the presence and velocities of radially inflowing molecular gas.

\section{Conclusions}\label{sec:conc}
The large field of view and sensitivity of the APEX maps offer a more complete picture of the molecular gas in Circinus and NGC~1097. In Circinus, the data helps confirm the presence of a molecular bar that was previously marginally detected, while in NGC~1097 the data reveals a previously unseen molecular gas component of the tidal tail. We summarize our key results as follows:
\begin{itemize}
    \item We present new APEX maps of the large-scale molecular gas of Circinus and NGC 1097, which are unprecedented at these sensitivity levels, and detect molecular gas at the largest extents so far for both galaxies, up to 5$^{\prime}$ (5 kpc) for Circinus, and $4^{\prime}.5$ (18 kpc) for NGC~1097.
    \item We compute molecular gas masses of $(2.1\times1.0)\times10^{9}$ M$_{\odot}$ for Circinus and $(4.7\pm1.8)\times10^{9}$ M$_{\odot}$ for NGC 1097, both of which are consistent with previous single-dish measurements of $M_{mol}$. The bright central regions in both galaxies comprise $\sim$65-75\% of the total luminosity of the whole galaxy. For Circinus, the ALMA 12-m data recovers $\sim35\%$ of the flux derived from the APEX data within the same region ($r\le1^{\prime}$, 1.2 kpc), while in the case of NGC~1097, the combined ALMA+ACA+TP flux is consistent with the APEX derived flux measurement.
    \item We detect new low-surface-brightness CO structures in both targets. Namely, in Circinus, we detect a molecular counterpart of a bar that had been only previously seen in the atomic gas and in thermal dust emission. In NGC 1097, we detect molecular gas which is likely being stripped away due to tidal interactions with the neighboring galaxy NGC 1097A.
    \item In Circinus, the large scale gas is found to rotate with a longer rotation period, while the gas within the inner $r\sim1$ kpc rotates in faster orbits. This results in an X-shape in the position-velocity diagram which is considered a diagnostic of a bar in edge-on disk galaxies. This result reaffirms previous indications of the presence of a bar in Circinus. The high curvature of the dust lane seen in the \textit{Spitzer} 8 $\mu$m continuum map points toward the bar being weak.
    \item NGC 1097 shows a tenuous (comprising $\sim2\%$  of the total molecular gas mass) tidal feature toward the direction of its companion galaxy NGC 1097A, an indication that a tidal interaction with the galaxy has led to the formation of molecular gas, or stripping away of gas into this region. We do not detect any shocked gas via the SiO(5--4) line in this region. The CO(2--1) emission peaks in the feature coincide with the location of peaks in the Spitzer 8$\mu$m continuum map, with a high velocity dispersion toward the brighter peak, and the feature exhibits a velocity gradient along its length.
    \item We attempt to fit models including radial flows to the gas kinematics of both galaxies using the tilted ring modeling software \bbarolo, but these results contain high uncertainties. It is therefore difficult to disentangle the rotation of gas in the disk from any non-axisymmetric flows such as radial flows along bars at the resolution and sensitivity of these maps beyond a simple analysis of the global kinematics with position-velocity diagrams. Nevertheless the data provide reasonably strict constraints on the geometry of the disk, i.e., position angle and inclination.
    \item We do not find any indications of optically dark companions in the molecular gas, or signatures of outflows or inflows in the APEX data for these galaxies.
\end{itemize}

This exploratory work with APEX shows the potential of future observations of extended and low-surface-brightness molecular gas components in and around nearby galaxies, to be conducted with higher sensitivity. Our study motivates the need for a new, large aperture, large FoV sub-millimeter single dish facility such as the proposed AtLAST \citep{2025A&A...694A.142M}. AtLAST's design includes a 50-m diameter primary mirror and a large instrument cabin that can simultaneously host up to six instruments that are quickly interchangeable, each with a FoV diameter of 1 or 2 degrees. Such a unique combination of collecting area and FoV results in an unprecedentedly high mapping speed of $\sim2.5\times$10$^{3}$ deg$^{2}$ mJy$^{-2}$ hr$^{-1}$ \citep{liu_atacama_2024}, which is more than four orders of magnitude larger than ALMA. Moreover, AtLAST will recover flux from structures as large as 1 degree. This would be ideal for mapping molecular emission lines tracing the extended cold ISM and CGM in and around nearby galaxies \citep{liu_atacama_2024, Lee+24}.

\begin{acknowledgements}
We thank the anonymous referee for their insightful suggestions, which has greatly improved the focus and quality of the manuscript. We would like to thank Enrico Di Teodoro for helpful discussions regarding $^{\rm 3D}$BAROLO. This project has received funding from the European Union's Horizon Europe and Horizon 2020 research and innovation programmes under grant agreements No. 101188037 (AtLAST2) and No. 951815 (AtLAST). Views and opinions expressed are however those of the author(s) only and do not necessarily reflect those of the European Union or European Research Executive Agency. Neither the European Union nor the European Research Executive Agency can be held responsible for them. The data was collected under the Atacama Pathfinder EXperiment (APEX) Project (ID: 0112.F-9507), led by the Max Planck Institute for Radio Astronomy at the ESO La Silla Paranal Observatory. This paper makes use of the following ALMA data: ADS/JAO.ALMA\#2017.1.00886.L, ADS/JAO.ALMA\#2013.1.00247.S. ALMA is a partnership of ESO (representing its member states), NSF (USA) and NINS (Japan), together with NRC (Canada), NSTC and ASIAA (Taiwan), and KASI (Republic of Korea), in cooperation with the Republic of Chile. The Joint ALMA Observatory is operated by ESO, AUI/NRAO and NAOJ. This research has made use of the NASA/IPAC Extragalactic Database (NED), which is funded by the National Aeronautics and Space Administration and operated by the California Institute of Technology.
\end{acknowledgements}

\bibliographystyle{aa}
\bibliography{bibliography}

\begin{appendix}
\section{Curve of growth}

\begin{figure}
\centering
\includegraphics[width=0.49\textwidth]{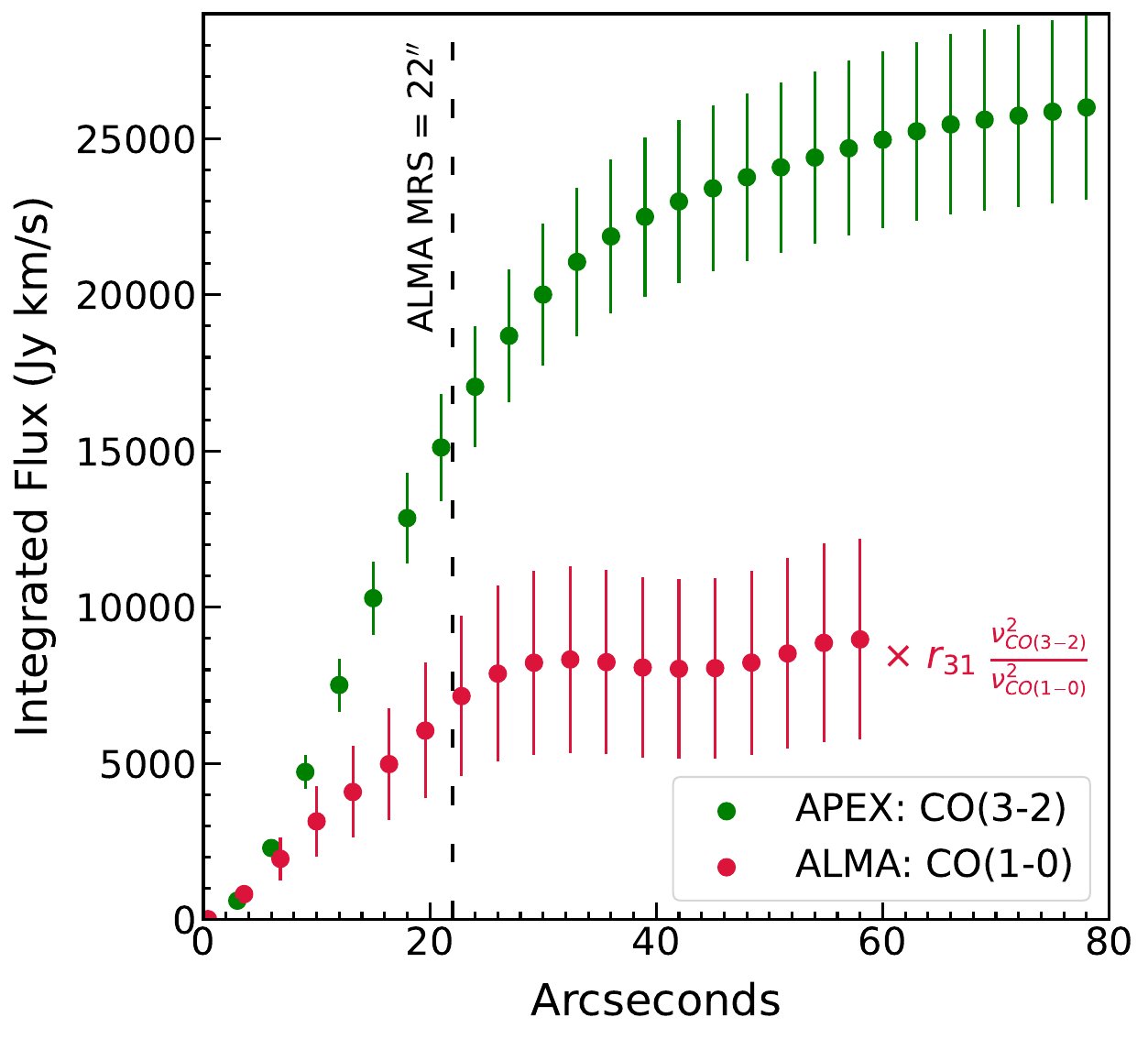}
\caption{Curve of flux growth in Circinus of APEX CO(3--2) and ALMA CO(1--0) (from \cite{2016ApJ...832..142Z}) emission within apertures of radius in increasing steps of $3.2^{\prime\prime}$ for ALMA, and $3^{\prime\prime}$ for APEX, respectively. The ALMA data was rescaled in flux density to CO(3--2) assuming a luminosity ratio of $r_{31} = 0.31\pm0.11$, and multiplying by the squared ratio of the rest frequencies of CO(3--2) ($\nu_{rest} = 345.796$ GHz) and CO(1--0) ($\nu_{rest} = 115.271$ GHz). The error bars for the ALMA data include a 5\% calibration uncertainty for Band 6 data, but a significant fraction of the error mainly arises from the assumption of the line luminosity ratio $r_{31}$, which is $\sim$35\%. The vertical dashed black line corresponds to the maximum recoverable scale of the ALMA data which is 22$^{\prime\prime}$, and is set by the shortest baseline length in the observing configuration.}
\label{fig:circ_cog}
\end{figure}

\begin{figure}
\centering
\includegraphics[width=0.49\textwidth]{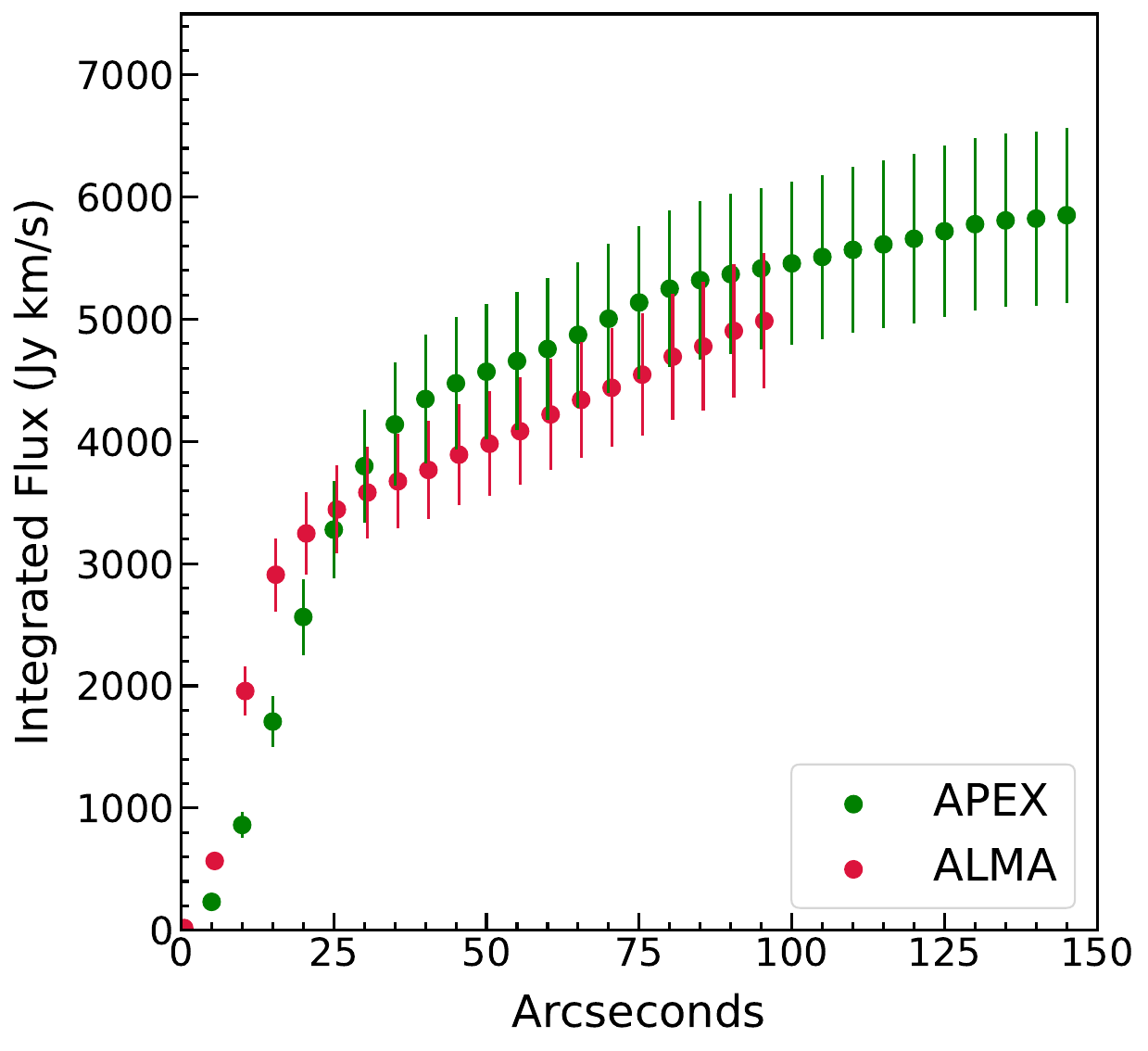}
\caption{Curve of flux growth in NGC 1097 of APEX CO(2--1) and ALMA CO(2--1) (from \cite{2021ApJS..257...43L}) emission within apertures in increasing steps of $5^{\prime\prime}$. The error bars for the ALMA data include (and are dominated by) a 10\% calibration uncertainty for Band 6 data.}
\label{fig:ngc_cog}
\end{figure}

In this section we plot the flux curve-of-growth of the APEX and ALMA archival data for Circinus (Figure \ref{fig:circ_cog}) and NGC~1097 (Figure \ref{fig:ngc_cog}), where the flux data points were calculated in concentric circular apertures of increasing radius around the center.

For Circinus, the APEX and ALMA flux levels are roughly consistent in $r<10^{\prime\prime}$, after which the increase in the ALMA flux is shallower. At $r>25^{\prime\prime}$ however, close to the maximum recoverable scale of the ALMA data, the ALMA flux values plateau while the APEX data continues to recover more flux and only increases at a shallower rate at $r>50^{\prime\prime}$ (which approximately corresponds to the radius of the bright central emission region).

In NGC~1097, the ALMA data are the combination of high resolution interferometric data with the Total Power (TP) array data which recovers the missing flux, and the consistency between the ALMA and APEX fluxes is showcased in the curve of growth.

\section{Channel maps}

\begin{figure*}
\centering
\includegraphics[width=\textwidth]{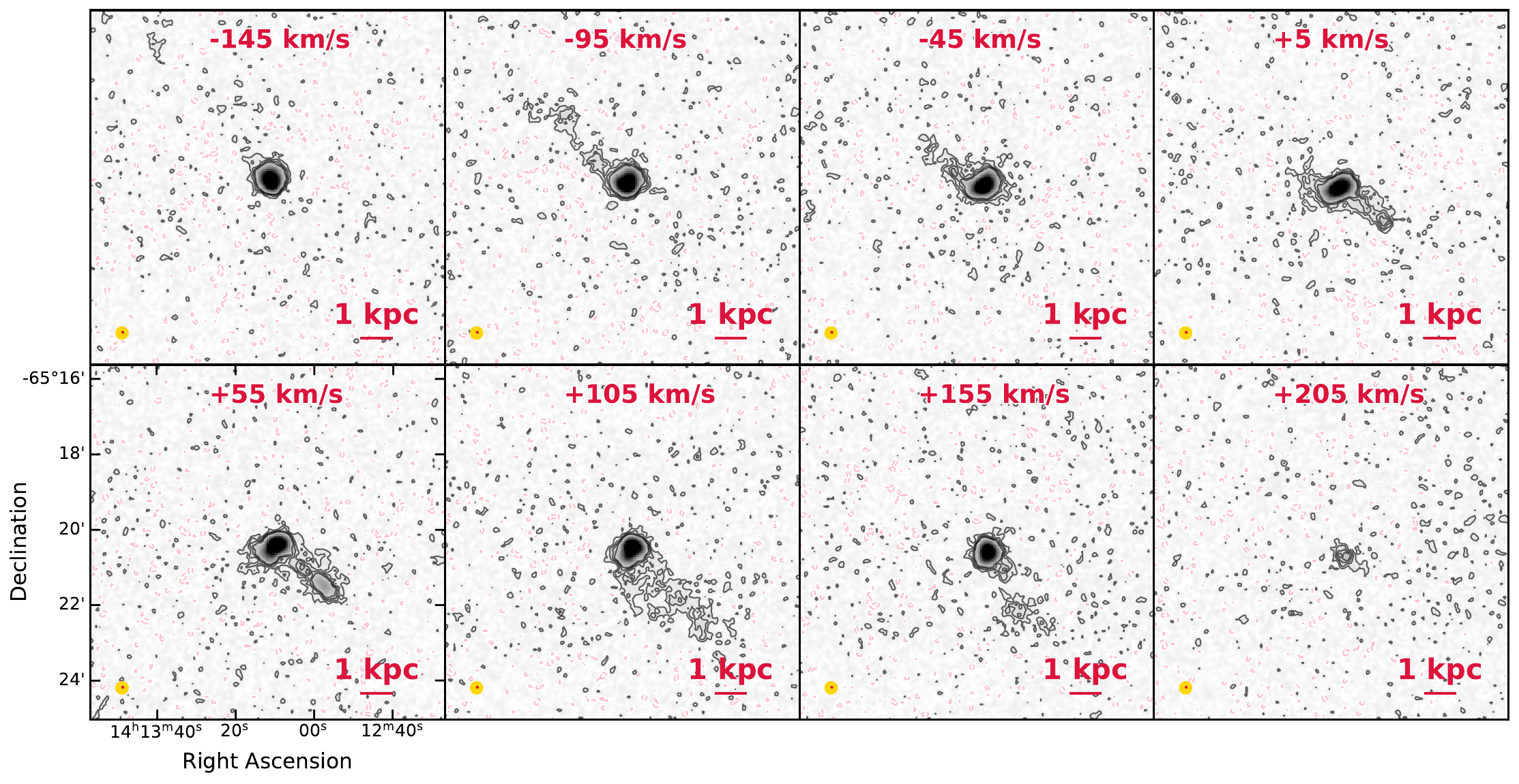}
\caption{Channel maps of the CO(3--2) emission in the Circinus galaxy, integrated over 10 channels (50 km/s). The channel maps are produced from the unmasked data. \textit{Contours:} The black contours show CO(3--2) emission contours in each channel at 2, 4, 6, 8 $\times~\sigma$, where $\sigma$ was calculated as the rms of the pixels in a circular annulus ($r_{in} = 225^{\prime\prime}$, $r_{out}= 285^{\prime\prime}$) excluding line emission from the galaxy in the center and noisy pixels toward the edges of the map. The pink contours show negative emission at a 2$\sigma$ level. The channels from +55 to +155 km/s show clumps of increased emission located about 1.8 kpc from the center. The beam sizes of the APEX (yellow) data are shown in the bottom left, and the bottom right scale bar marks a projected distance of 1 kpc.}
\label{fig:circ_chanmap}
\end{figure*}

\begin{figure*}
\centering
\includegraphics[width=\textwidth]{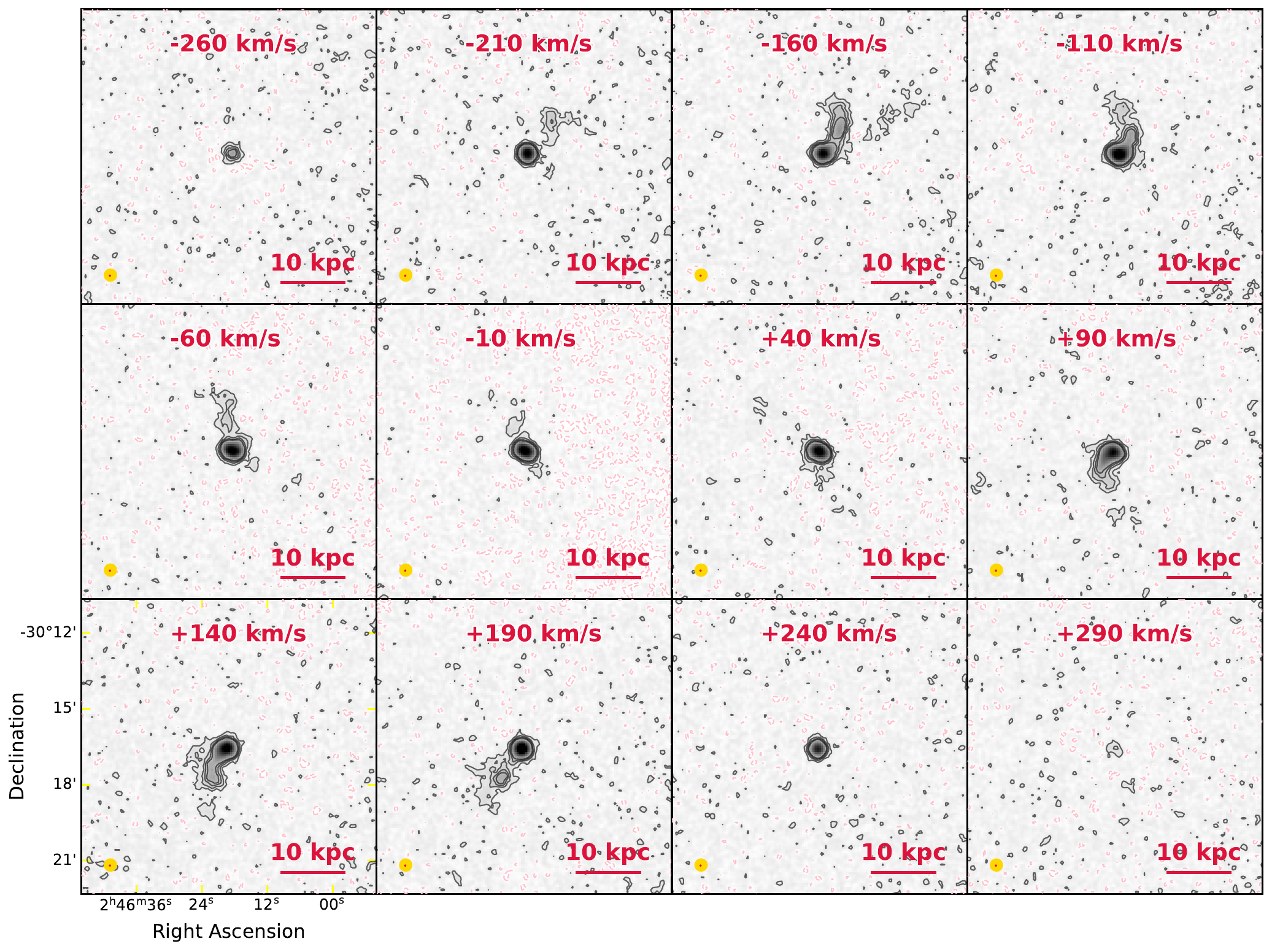}
\caption{Channel maps of the CO(2--1) emission in NGC 1097, integrated over 10 channels (50 km/s). The channel maps are produced from the unmasked data. \textit{Contours:} The black contours show CO(2--1) emission contours in each channel at 2, 5, 8, 11 $\times~\sigma$, where $\sigma$ was calculated from a circular annulus ($r_{in} = 250^{\prime\prime}$, $r_{out} = 350^{\prime\prime}$) excluding pixels with line emission and noisy pixels at the edges of the map. Pink contours show negative emission features at a 2$\sigma$ level. The beam sizes of the APEX (yellow) data are shown in the bottom left, and the bottom right scale bar marks a projected distance of 10 kpc.}
\label{fig:ngc_chanmap}
\end{figure*}

For visualizing the gas distribution at different velocities, we plot successive channel maps of the CO(3--2) and CO(2--1) emission, integrated over 10 channels (50 km/s) each, for Circinus and NGC~1097 in Figures \ref{fig:circ_chanmap} and \ref{fig:ngc_chanmap}, respectively.

For Circinus, we see the furthest detections of gas at $v = -145$ km/s, where we detect emission at 2$\sigma$ levels at a distance of $\sim$4.8$^{\prime}$ ($\sim$5.9 kpc) away. In the receding gas distribution, the furthest gas detected is seen in the clumpy features at $v=+105$ km/s at a distance of $\sim$3$^{\prime}$ ($\sim$3.6 kpc).

In the case of NGC~1097, we most notably see the tidal feature in the $v = -160$ km/s channel, which is furthest detection of gas in the map, at about 4.5$^{\prime}$ ($\sim$18 kpc) from the center. Furthermore, at $v = +140$ km/s, there also appears to be a disconnected clump of diffuse emission at a 2$\sigma$ level toward the south, although this is a less robust detection. The furthest detection toward this directions is seen at $v = +190$ km/s, where the gas is seen at an extent of 3$^{\prime}$ ($\sim$14 kpc).

\section{\bbarolo model maps}

\begin{figure*}
\centering
\includegraphics[width=1\textwidth]{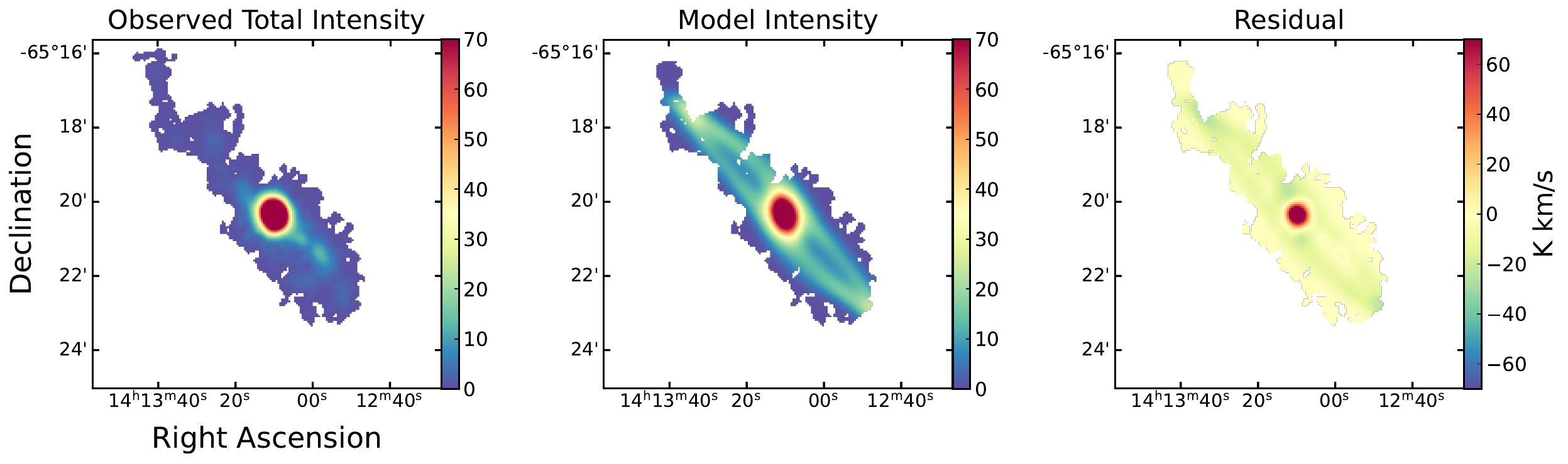}
\includegraphics[width=1\textwidth]{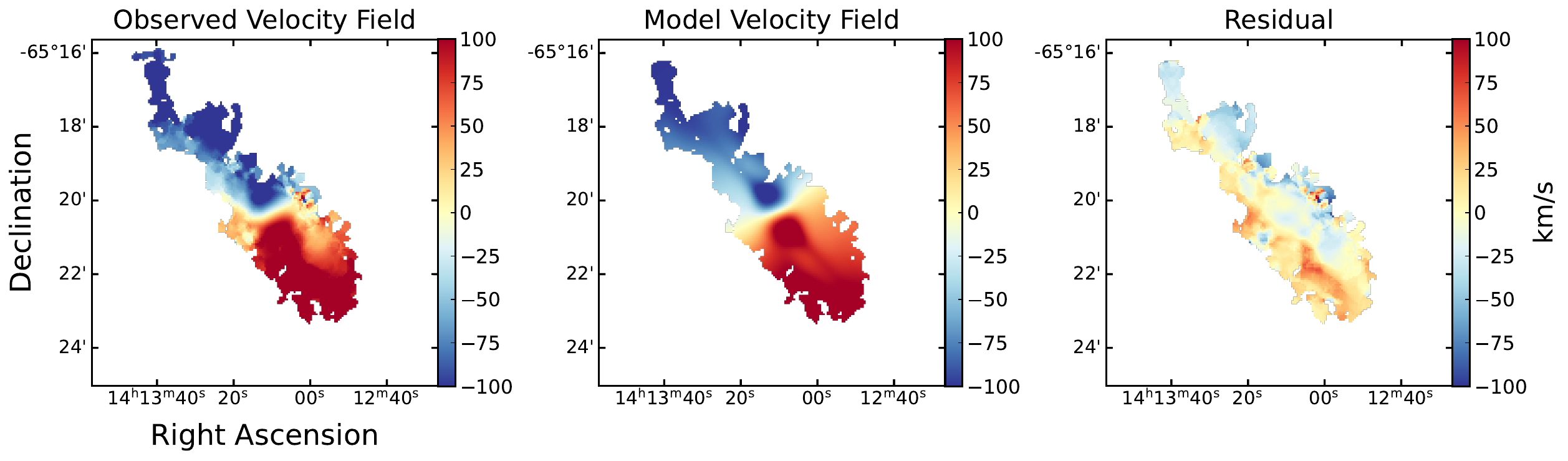}
\includegraphics[width=1\textwidth]{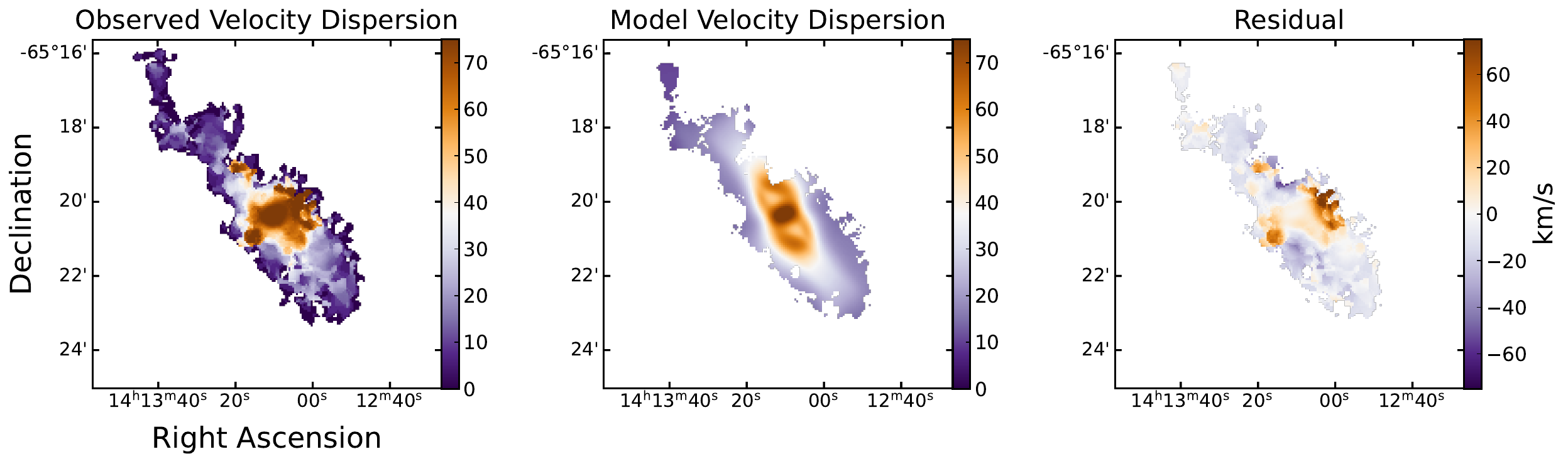}
\caption{Results from the 3D tilted ring modeling of Circinus in \bbarolo fit without a radial velocity component. The first column shows the integrated intensity map, velocity field, and dispersion map computed by \bbarolo from the data, the second column shows the modeled integrated intensity, velocity field, and dispersion map, and the final column shows the respective residuals (Data -- Model). The color maps are fixed to the same range in each row for a straightforward comparison with the residuals.}
\label{fig:circinus_bb_novrad}
\end{figure*}

\begin{figure*}
\centering
\includegraphics[width=1\textwidth]{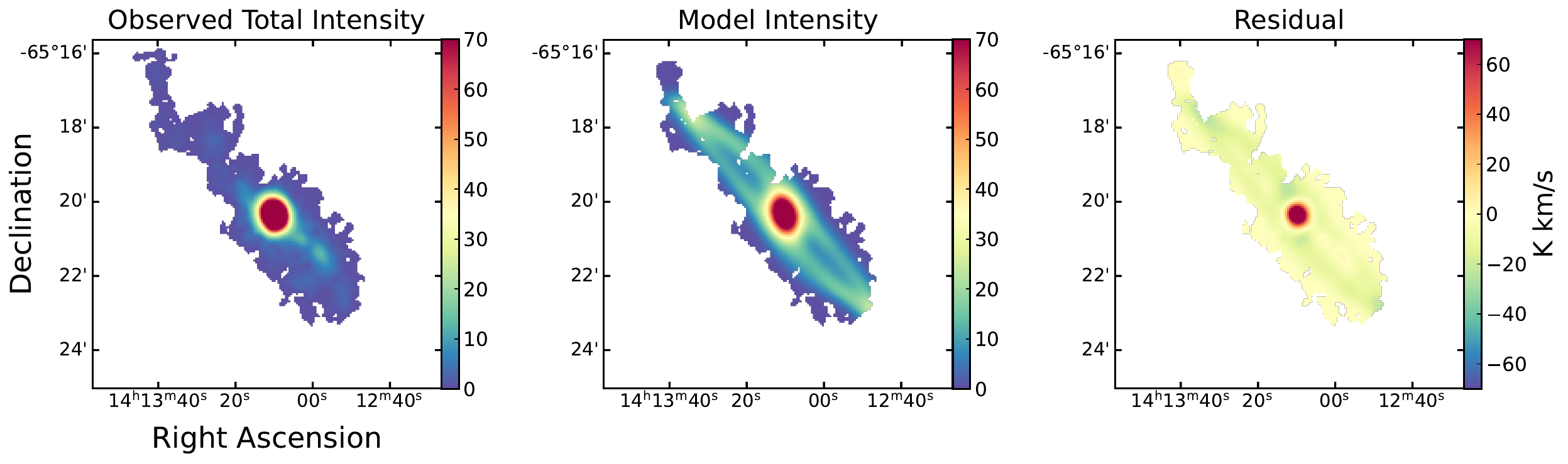}
\includegraphics[width=1\textwidth]{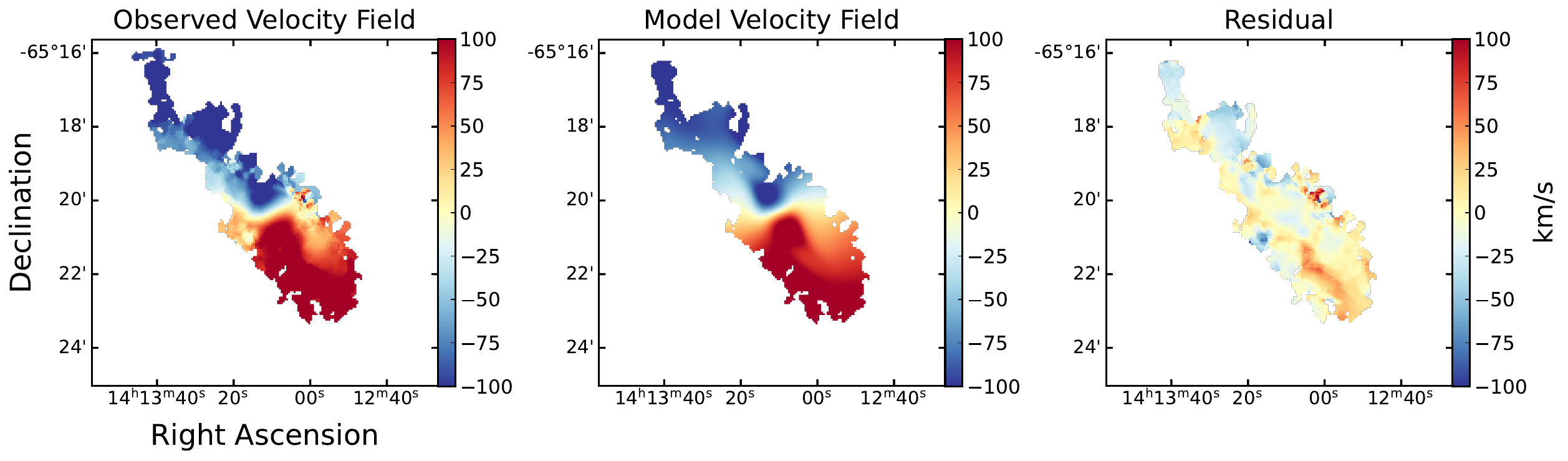}
\includegraphics[width=1\textwidth]{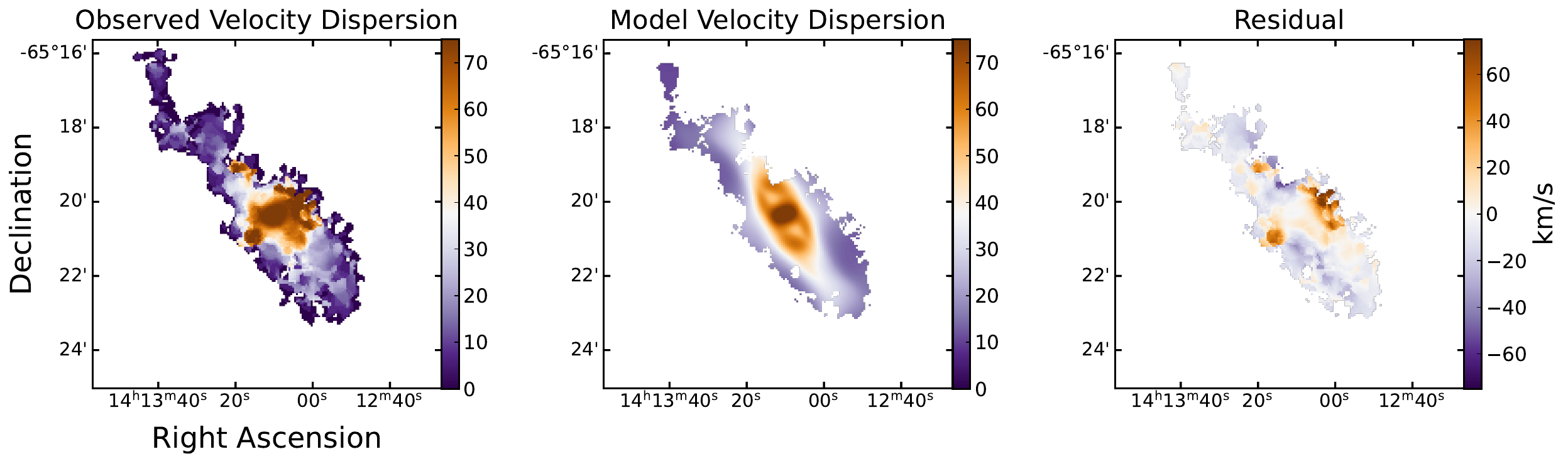}
\caption{Results from the 3D tilted ring modeling of Circinus in \bbarolo fit including a radial velocity component. The first column shows the integrated intensity map, velocity field, and dispersion map computed by \bbarolo from the data, the second column shows the modeled integrated intensity, velocity field, and dispersion map, and the final column shows the respective residuals (Data -- Model). The color maps are fixed to the same range in each row for a straightforward comparison with the residuals.}
\label{fig:circinus_bb_vrad}
\end{figure*}

\begin{figure*}
\centering
\includegraphics[width=1\textwidth]{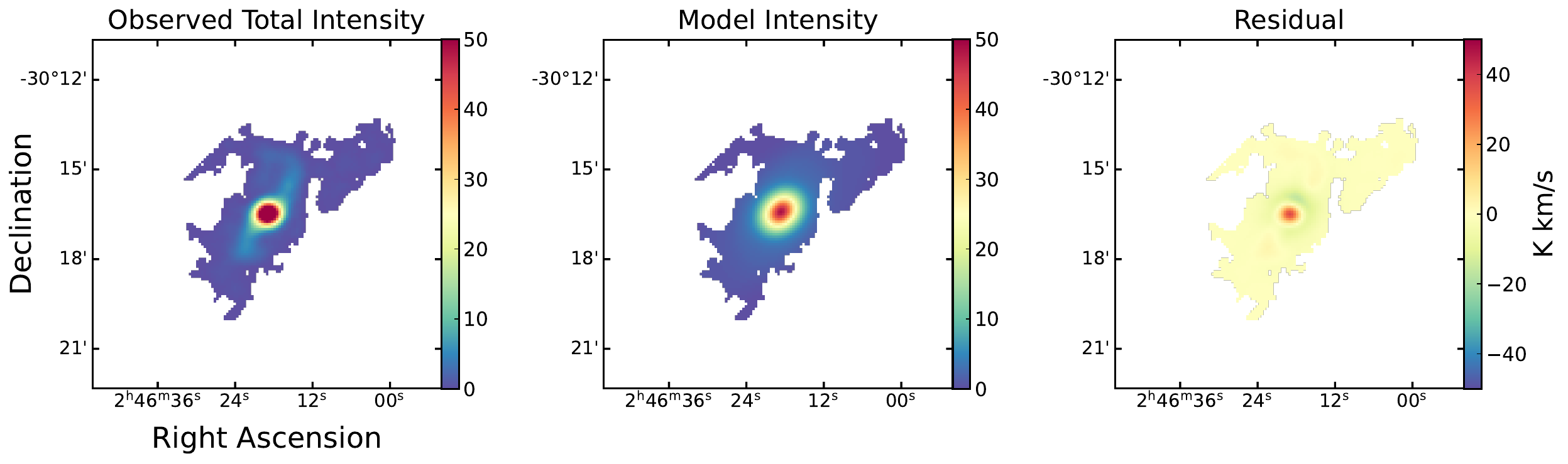}
\includegraphics[width=1\textwidth]{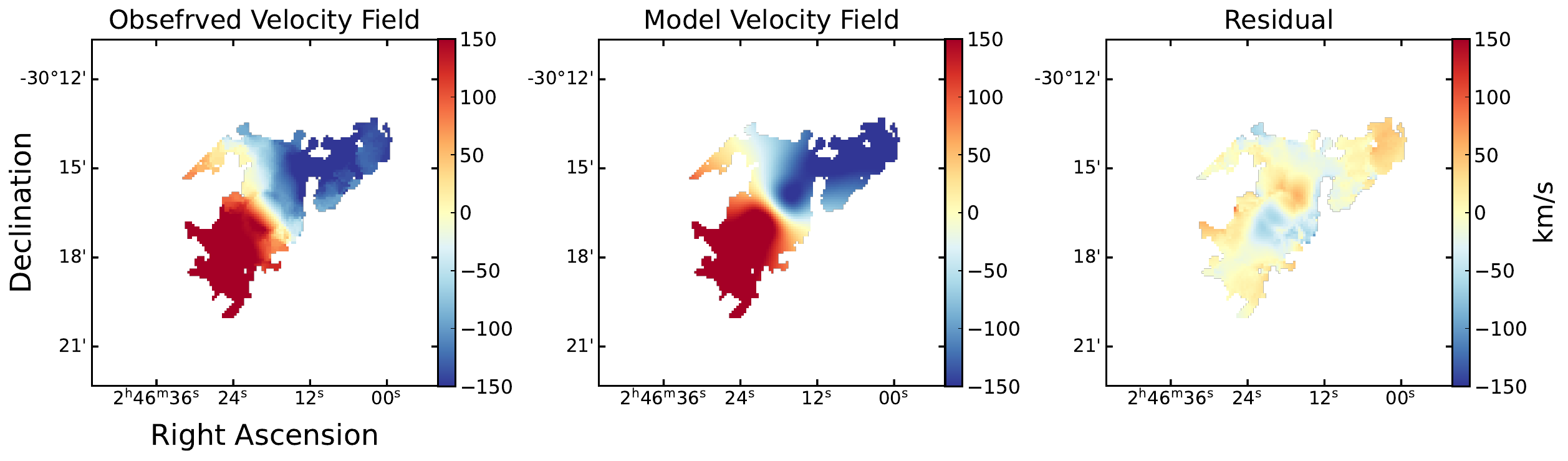}
\includegraphics[width=1\textwidth]{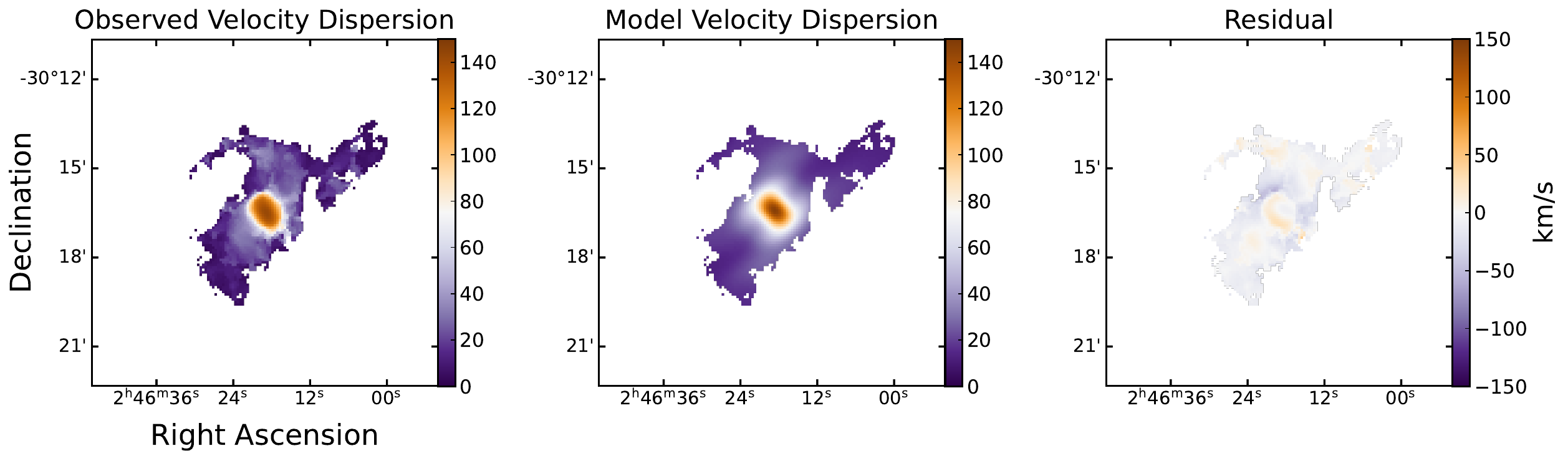}
\caption{Results from the 3D tilted ring modeling of NGC 1097 in \bbarolo fit without a radial velocity component. The first column shows the integrated intensity map, velocity field, and dispersion map computed by \bbarolo from the data, the second column shows the modeled integrated intensity, velocity field, and dispersion map, and the final column shows the respective residuals (Data -- Model). The color maps are fixed to the same range in each row for a straightforward comparison with the residuals.}
\label{fig:ngc1097_bb_novrad}
\end{figure*}

\begin{figure*}
\centering
\includegraphics[width=1\textwidth]{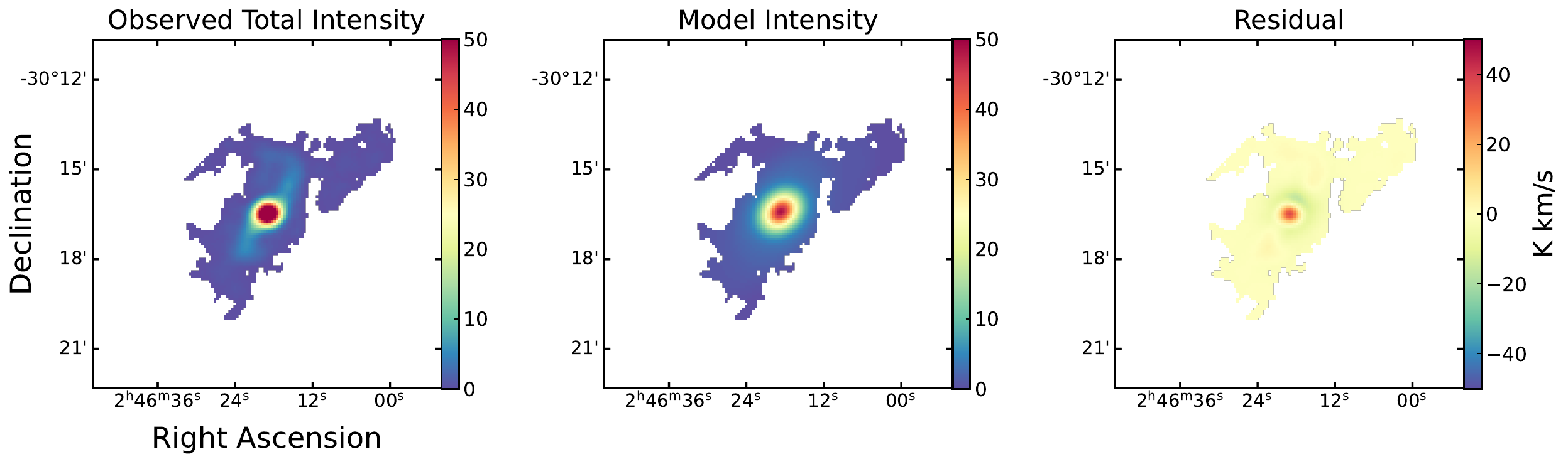}
\includegraphics[width=1\textwidth]{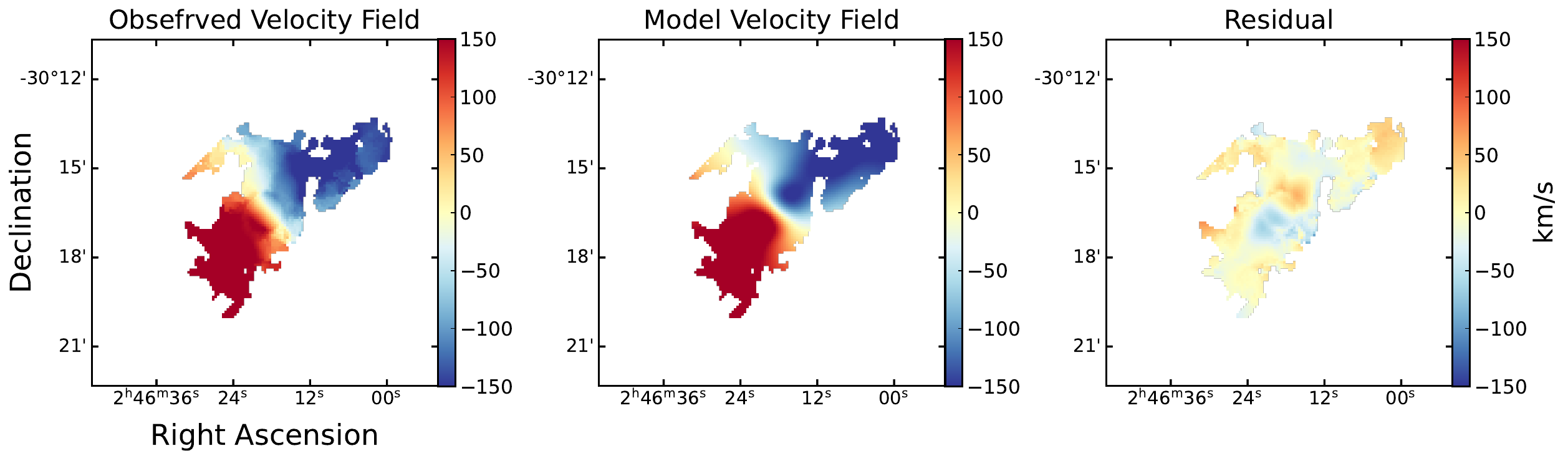}
\includegraphics[width=1\textwidth]{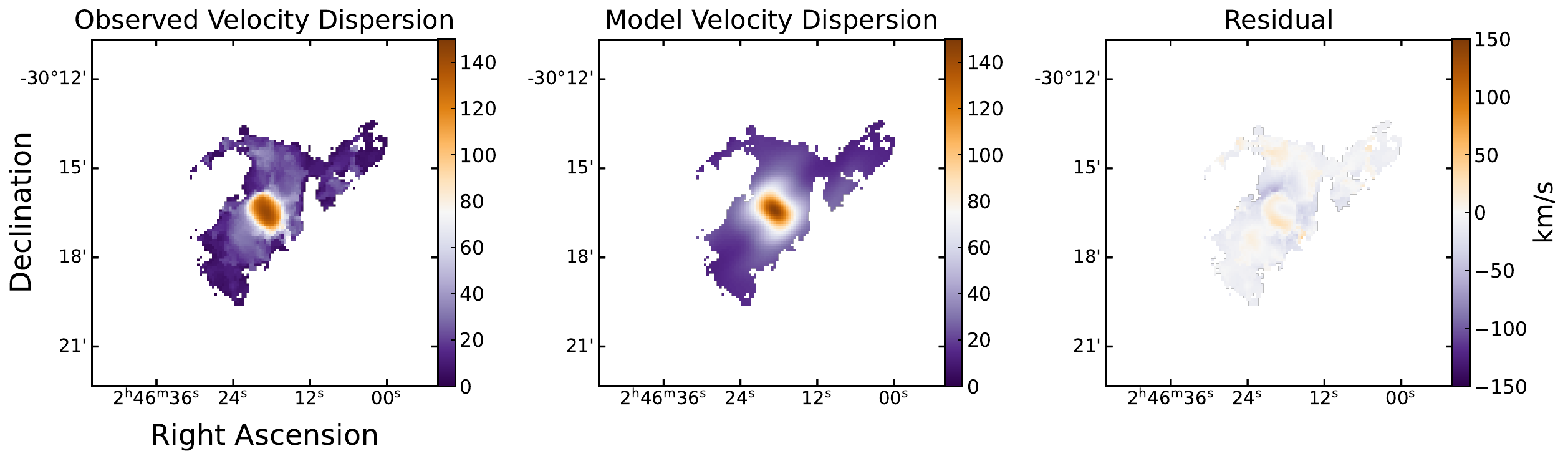}
\caption{Results from the 3D tilted ring modeling of NGC 1097 in \bbarolo fit including a radial velocity component. The first column shows the integrated intensity map, velocity field, and dispersion map computed by \bbarolo from the data, the second column shows the modeled integrated intensity, velocity field, and dispersion map, and the final column shows the respective residuals (Data -- Model). The color maps are fixed to the same range in each row for a straightforward comparison with the residuals.}
\label{fig:ngc1097_bb_vrad}
\end{figure*}

In this section we present plots of the integrated intensity, velocity centroid, and velocity dispersion maps for the \bbarolo models of Circinus and NGC~1097, without and with the inclusion of a radial velocity component, discussed in \S \ref{sec:circ_kin_bbar}, \ref{sec:ngc_kin_bbar}, and \ref{sec:disc_tilted_modeling}. Figures \ref{fig:circinus_bb_novrad} and \ref{fig:circinus_bb_vrad} show the corresponding moment maps for the Circinus galaxy, while Figures \ref{fig:ngc1097_bb_novrad} and \ref{fig:ngc1097_bb_vrad} show the maps for NGC~1097.

\end{appendix}
\end{document}